\documentclass[aps,prd,onecolumn,groupedaddress,nofootinbib,amssymb]{revtex4}  
\usepackage[T1]{fontenc}
\usepackage[latin1]{inputenc}
\usepackage{graphicx}
\usepackage[english]{babel}

\begin{document}

\title{Relativistic orbits with gravitomagnetic corrections}

\author{S. Capozziello$^1$, M. De Laurentis$^{1,2}$, F. Garufi$^1$, and L. Milano$^1$}

\affiliation{\it $^1$Dipartimento di Scienze fisiche, Università
di Napoli {}`` Federico II'' and INFN Sez. di Napoli, Compl. Univ.
di
Monte S. Angelo, Edificio G, Via Cinthia, I-80126, Napoli, Italy\\
$^2$Dipartimento di Fisica, Politecnico di Torino and INFN Sez. di
Torino, Corso Duca degli Abruzzi 24, I-10129 Torino, Italy}
\date{\today}

\begin{abstract}
Corrections to the relativistic  orbits are  studied considering
higher order approximations induced by gravitomagnetic effects. We
discuss in details how such corrections come out taking into
account "magnetic"  components in the weak field limit  of
gravitational field and then the theory of orbits is developed
starting from the Newtonian one, the lowest order in the
approximation. Finally, the orbital structure and the stability
conditions are discussed giving numerical examples. Beside the
standard periastron corrections of General Relativity, a  new
nutation effect is due to the ${\displaystyle c^{-3}}$
corrections. The transition to a chaotic behavior strictly depends
on the initial conditions. The orbital phase space portrait is
discussed.

\end{abstract}
\maketitle


\vspace{4. mm}

\section{Introduction}
The  analogy between the classical Newton and Coulomb laws led to
investigate if  masses in motion, considered as charges, could
give rise to a "gravitational" magnetic field.

In fact, the magnetic field is produced by the motion of
electric-charge, i.e. the electric current. The analogy consists
of the fact that a mass-energy current can produce what is called
"gravitomagnetic" field.

The pioneering approach to the problem is due to Maxwell himself
which, in one of his fundamental works on electromagnetism, turned
his attention on the possibility to formulate the theory of
gravitation in a form corresponding to the electromagnetic
equations \cite{Maxwell}. However, he was puzzled by the problem
of the energy of the gravitational field, i.e. the meaning and the
origin of the negative energy due to the mutual attraction of
material bodies. In fact, according to him, the energy of a given
field had to be "essentially positive", but this is not the case
for the gravitational field. To balance this negative energy, a
great amount of positive energy is required, in the form of energy
of the space (a sort of back-reaction). But, since he was unable
to understand how this could be, he did not proceed further along
this line of thinking since the problem can be addressed and
solved only in the framework of General Relativity.

Later, Holzmuller \cite{Holzmuller} and Tisserand \cite{Tisserand}
proposed to modify the Newton law  introducing, in the radial
component of the force,  a term depending on the relative velocity
of the two attracting particles (see also \cite{North,Whittaker}).
Also Heaviside \cite{Heaviside,Heaviside2} investigated the
analogy between gravitation and electromagnetism considering the
propagation of gravitational energy in terms of a sort of gravito
- electromagnetic Poynting vector: however,  also in this case, he
failed to frame the problem of gravitational energy in a
self-consistent scheme.

Finally, the formal analogy between electromagnetic and
gravitational fields was explored by Einstein \cite{Einstein}, in
the framework of General Relativity, and then by Thirring
\cite{Thirring1}. This author pointed out that the geodesic
equation can be written as a Lorentz force splitting the
gravitational field in gravito-electric and gravito-magnetic
components. The final result of these studies was that any theory
which combines Newtonian gravity together with Lorentz invariance
in a consistent way, has to include a gravitomagnetic field, which
is generated by the mass-energy current. This is the case, of
course, of General Relativity: it was shown by Lense and Thirring
\cite{Thirring}, that a rotating mass generates a gravitomagnetic
field, which in turn, causes the precession of planetary orbits.
To be more precise, H. Pfister has recently shown that it would be
better to speak about an Einstein - Lense - Thirring effect
\cite{pfister}.

It is  interesting to notice that also Lodge and Larmor, at the
end of the nineteenth century, discussed the effects of frame
dragging on a non-rotating interferometer \cite{Anderson}, but
within the framework of an aether-theoretical model. This frame
dragging corresponded, in fact, to the Lense-Thirring effect of
General Relativity. However, at the beginning of the XX century,
when Lense and Thirring published their  papers, the effect named
after them, which is indeed very small in the terrestrial
environment, was far from being detectable, because of the
technical difficulties and limitations of the time. Contemporary
improvements in technology have made possible to propose new ideas
to reveal the Lense-Thirring precession by analyzing the data-sets
on the orbits of Earth satellites (see e.g. \cite{cugusi} where,
for the first time, the use of LAGEOS satellite was proposed).
Several proposals have been recently published to measure the
Lense-Thirring effect by natural and artificial bodies in some
Solar System scenarios. For example, in \cite{iorio1}, the Sun
with Venus is considered. Mars with MGS spacecraft is discussed in
\cite{iorio2}, while Jupiter with the Galilean moons (which is the
original idea by Lense and Thirring) is studied in \cite{iorio3}.
Regarding the Earth with the existing LAGEOS and LAGEOSII
satellites, recent results are reported in \cite{ries}, while for
the approved experiment LARES the expected forthcoming measurement
are discussed in \cite{iorio4}.

On the other hand, the  experiment Gravity Probe-B \cite{Everitt}
has been devoted to another gravitomagnetic effect due to Earth's
rotation, i.e. the Pugh-Schiff effect consisting of the
precessions of the spins of four gyroscopes carried onboard the
spcaecraft \cite{pugh,schiff}. This experiment has detected the
effect and its magnitude in the gravitational field of the Earth
\cite{gpb}. The originally expected accuracy was $1\%$ or better,
but it is still unclear if it will be finally obtained because of
unexpected systematic effects arisen in the data analysis.  Other
experiments (like GP-C \cite{Gronwald,iorio5,iorio6}) have been
proposed to reveal the space-time structure, which is affected by
gravitomagnetism, for example evidencing clock effects around a
spinning massive object. In particular, concerning the so-called
gravitomagnetic clock effect, we have to stress that its most
investigated form consists of the difference between the orbital
periods of two counter-rotating satellites.

Recently, gravitomagnetic effects have been considered also in the
framework of gravitational lensing. By using the Fermat principle
and the standard theory of gravitational lensing, the
gravitomagnetic corrections to the time delay function and the
deflection angle for a geometrically thin lens can be derived.
Such corrections can induce observational effects both in
point-like \cite{capozzlamb} and in extended gravitational lenses
(as the isothermal sphere and the disk of spiral galaxies
\cite{capozzre,sereno}. Other researches concerning the
gravitomagnetic effects on time delay and light deflection have
been pursued. In \cite{kopeikin}, the gravitomagnetic effects in
the propagation of electromagnetic waves in variable gravitational
fields of arbitrary-moving and spinning bodies have been studied,
while, in \cite{sereno1,sereno2,sereno3}, the gravitational
lensing due to stars with angular momentum, and then inducing
gravitomagnetic effects,  have been considered.

Finally, the analogy between general relativity and
electromagnetism suggests that there is also a
galvano-gravitomagnetic effect, which is the gravitational analog
of the Hall effect. This  effect takes place when a current
carrying conductor is placed in a gravitomagnetic field and the
conduction electrons moving inside the conductor are deflected
transversally with respect to the current flow. Such a
galvano-gravitomagnetic effect, considering current carrying
conductors, could be used for detecting the gravitomagnetic field
of the Earth. A discussion of the effect and its measurability is
in \cite{ahmedov1,ahmedov2,iorio7}.

In this paper, we want to study how the relativistic theory of
orbits for massive point-like  objects is affected by
gravitomagnetic corrections. In other words, we want to consider
the orbital effects of higher-order terms in $v/c$ and this is the
main difference with respect to the standard gravitomagnetic
effect so far considered. In this case, the problem of
gravitomagnetic vector potential entering into the off-diagonal
components $g_{0l}$ of the metric $g_{\mu\nu}$ can be greatly
simplified and the corrections can be seen as  further powers in
the expansion in $c^{-1}$ (up to $c^{-3}$). Nevertheless, the
effects on the orbit behavior are interesting and involve not only
the precession at peri-astron but also nutation corrections as we
will show below. This means that it could be misleading to neglect
such effects when the weak field approximation is not so weak, as
in the case of point-like compact objects moving in a
tight-binding regime or spiralizing each other as in the case of
evolved binary  systems constituted by black holes and/or neutron
stars. A study in this sense is in \cite{iorio8} where the
possibility of measuring the Lense-Thirring effect with the double
pulsar J0737-3039A is discussed.

In particular, we can study the evolution of compact binary
systems in the extreme mass ratio limit, i.e.  the mass of the
moving particle is $m$ and the mass that produces the
gravitational field is $M$, so that $m \ll M$. This constraint is
satisfied by several real systems. For example, there has been
gathering evidence suggesting the existence of supermassive black
hole with masses in the ranges $10^{6}\div 10^{9}M_{\odot}$) in
galactic nuclei \cite{Wald,Kormendy}. One expects that small
compact objects ($1\div10 M_{\odot}$) from the surrounding stellar
population will be captured by these black holes following
many-body scattering interactions at a relatively high rate
\cite{Sigurdsson,Sigurdsson2}.

Our approach  suggests that, in the weak field approximation, when
considering higher order corrections in the motion equations, the
gravitomagnetic effects can be particularly significant, also in a
rough approximation, giving rise also to  chaotic behaviors in the
transient regime dividing stable from unstable orbits. Generally,
such contributions are discarded since they are assumed too small
 but they have to be taken into account as soon as the $v/c$ is
not so small.

Sec.II is devoted to the discussion of the gravitomagnetic
corrections which have to be considered when relevant mass-energy
current effects are presented in a given problem. The geodesics,
and then their spatial components,  the trajectories, are
corrected by such terms. We derive the Christoffel symbols with
gravitomagnetic corrections and the vector form of geodesics. In
particular,  the metric "gravitomagnetically" corrected is
achieved and the conditions in which the vector potential $V^l$
can be substituted with its point-like counterpart $\Phi v^l/c$
where $\Phi$ is the static Newton potential and $v^l$ the velocity
of the test-particle $m$ moving around the generator of the
gravitational field $M$.

In Sec. III, the theory of orbits is discussed.  We review the
Newtonian and the relativistic theory considering, in particular,
the role of relativistic corrections \cite{binney,landau}.   In
Sec.IV, after constructing an effective Lagrangian coming from the
line element with the gravitomagnetic effect,  we derive the
equations of motion. Numerical results for orbits and their
phase-space portrait are presented in Sec.V. Discussion and
conclusions are drawn in Sec.VI.

\section{Gravitomagnetic effects}

Before treating the theory of the orbits with the gravitomagnetic
effects, let us get  some insight into  gravitomagnetism and show
how to derive the corrected metric. A recent book concerning both
theoretical and experimental aspects of gravitomagnetism is
\cite{iorio9}, while the Lense-Thirring effect is discussed in
\cite{ruffini}.

A remark is in order at this point: any theory  combining, in a
consistent way,  Newtonian gravity together with Lorentz
invariance  has to include a gravitomagnetic field generated by
the mass-energy currents. This is the case, of course, of General
Relativity: it was shown by Lense and Thirring
\cite{Thirring,barker,ashby,iorio10,Tartaglia}, that a rotating
mass generates a gravitomagnetic field, which, in turn, causes a
precession of planetary orbits. In the framework of the linearized
weak-field and slow-motion approximation of General Relativity,
the ensemble of the so-called gravitomagnetic effects are induced
by the off-diagonal components of the space-time metric tensor
which are proportional to the components of the matter-energy
current density of the source. It is possible to take into account
two types of mass-energy currents in gravity. The former is
induced by the matter source rotation around its center of mass:
it generates the intrinsic gravitomagnetic field which is closely
related to the angular momentum (spin) of the rotating body. The
latter is due to the translational motion of the source: it is
responsible of the extrinsic gravitomagnetic field. This concept
has been discussed in Refs.\cite{pascual,kopeikin2}. Then,
starting from the Einstein field equations in the weak field
approximation one obtain the gravitoelectromagnetic equations and
then the corrections in the metric.  Let us start from the weak
field approximation of the gravitational field\footnote{Notation:
latin indices run from 1 to 3, while greek indices run from 0 to
3; the flat spacetime metric tensor is
$\eta_{\mu\nu}=diag(1,-1,-1,-1)$.}

\begin{equation}
g_{\mu\nu}(x)=\eta_{\mu\nu}+h_{\mu\nu}(x),\qquad\left|h_{\mu\nu}(x)\right|<<1.\label{eq:g_muni}\end{equation}
where $\eta_{\mu\nu}$ is the Minkowski metric tensor and
$\left|h_{\mu\nu}(x)\right|<<1$ is a small deviation from it
\cite{weinberg}.

The stress-energy tensor for perfect - fluid matter is given by

\begin{equation}
T^{\mu\nu}=\left(p+\rho
c^{2}\right)u^{\mu}u^{\nu}-pg^{\mu\nu}\end{equation}
which, in the
weak field approximation $p\ll\rho c^{2}$, is

\begin{equation}
T^{00}\simeq\rho c^{2},\qquad T^{0j}\simeq\rho cv^{j},\qquad
T^{ij}\simeq\rho v^{i}v^{j}\,.\end{equation} From the Einstein
field equations $G_{\mu\nu}=(8\pi G/c^4)T_{\mu\nu}$, one finds

\begin{equation}
\bigtriangledown^{2}h_{00}=\frac{8\pi
G}{c^{2}}\rho\,,\label{eq:nabla_00}\end{equation}

\begin{equation}
\bigtriangledown^{2}h_{ij}=\frac{8\pi
G}{c^{2}}\delta_{ij}\rho\,,\label{eq:nabla_ij}\end{equation}

\begin{equation}
\bigtriangledown^{2}h_{0j}=-\frac{16\pi G}{c^{2}}\delta_{jl}\rho
v^{l}\,,\label{eq:nabla_0j}\end{equation}

where $\bigtriangledown^{2}$ is the standard Laplacian operator
defined on the flat spacetime. To achieve Eqs.
(\ref{eq:nabla_00})-(\ref{eq:nabla_0j}),  the harmonic condition

\begin{equation}
g^{\mu\nu}\Gamma_{\mu\nu}^{\alpha}=0\;,\end{equation} has been
used.

By integrating  Eqs. (\ref{eq:nabla_00})-(\ref{eq:nabla_0j}), one
obtains

\begin{equation}
h_{00}=-\frac{2\Phi}{c^{2}}\;,\label{eq:h_00}\end{equation}

\begin{equation}
h_{ij}=-\frac{2\Phi}{c^{2}}\delta_{ij}\;,\label{eq:h_ij}\end{equation}

\begin{equation}
h_{0j}=\frac{4}{c^{3}}\delta_{jl}V^l\;.\label{eq:h_0j}\end{equation}
The metric is determined by the gravitational Newtonian potential

\begin{equation}
\Phi(x)=-G\int\frac{\rho}{\left|\mathbf{x}-\mathbf{x}'\right|}d^{3}x'\;,\label{eq:fi_x}\end{equation}

and by the vector potential $V^{l}$,

\begin{equation}
V^{l}=-G\int\frac{\rho
v^{l}}{\left|\mathbf{x}-\mathbf{x}'\right|}d^{3}x'\;.\label{eq:Vl}\end{equation}
given by the  matter current density $\rho v^{l}$ of the moving
bodies. This last potential gives rise to the gravitomagnetic
corrections.

From Eqs(\ref{eq:g_muni}) and (\ref{eq:h_00})-(\ref{eq:Vl}),  the
metric tensor in terms of Newton and gravitomagnetic potentials is

\begin{equation}
ds^{2}=\left(1+\frac{2\Phi}{c^{2}}\right)c^{2}dt^{2}-
\frac{8\delta_{lj}V^{l}}{c^{3}}cdtdx^{j}-\left(1-\frac{2\Phi}{c^{2}}\right)\delta_{lj}dx^{i}dx^{j}\;.
\label{eq:ds_DUE}\end{equation}

From Eq.(\ref{eq:ds_DUE}) it is possible to construct a
variational principle from which the geodesic equation follows.
Then we can derive the orbital equations. As standard, we have

\begin{equation}
\ddot{x}^{\alpha}+\Gamma_{\mu\nu}^{\alpha}\dot{x}^{\mu}\dot{x}^{\nu}=0\;,\label{eq:geodedica_uno}\end{equation}

where the dot indicates the differentiation with respect to the
affine parameter. In order to put in evidence the gravitomagnetic
contributions, let us explicitly calculate the Christoffel symbols
at lower orders. By some straightforward calculations, one gets
\begin{equation}\begin{array}{cl}
\Gamma^0_{00} &=0\\
\Gamma^0_{0j} &=\frac{1}{c^2}\frac{\partial\Phi}{\partial x^j} \\
\Gamma^0_{ij} &=-\frac{2}{c^3}\left(\frac{\partial V^i}{\partial x^j}+\frac{\partial V^j}{\partial x^i}\right) \\
\Gamma^k_{00} &= \frac{1}{c^2}\frac{\partial\Phi}{\partial x^k}\\
\Gamma^k_{0j} &=\frac{2}{c^3}\left(\frac{\partial V^k}{\partial x^j}-\frac{\partial V^j}{\partial x^k}\right) \\
\Gamma^k_{ij} &= -\frac{1}{c^2}\left(\frac{\partial \Phi}{\partial
x^j}\delta^k_i+\frac{\partial \Phi}{\partial
x^i}\delta^k_j-\frac{\partial \Phi}{\partial
x^k}\delta_{ij}\right)\end{array}\end{equation} In the
approximation which we are going to consider, we are retaining
terms up to the orders $\Phi/c^2$ and $V^j/c^3$. It is important
to point out that we are discarding terms like
$(\Phi/c^4)\partial\Phi/\partial x^k$,
$(V^j/c^5)\partial\Phi/\partial x^k$, $(\Phi/c^5)\partial
V^k/\partial x^j$, $(V^k/c^6)\partial V^j/\partial x^i$ and of
higher orders. Our aim is to show that, in several  cases like in
tight binary stars, it is not correct to discard higher order
terms in $v/c$ since physically interesting effects could come
out.

The geodesic equations up to $c^{-3}$ corrections are then

\begin{equation}
c^{2}\frac{d^{2}t}{d\sigma^{2}}+\frac{2}{c^{2}}\frac{\partial\Phi}{\partial
x^{j}}c\frac{dt}{d\sigma}\frac{dx^{j}}{d\sigma}-\frac{2}{c^{3}}\left(\delta_{im}\frac{\partial
V^{m}}{\partial x^{j}}+\delta_{jm}\frac{\partial V^{m}}{\partial
x^{i}}\right)\frac{dx^{i}}{d\sigma}\frac{dx^{j}}{d\sigma}=0\;,\label{time}
\end{equation}
for the time component, and

\begin{eqnarray}
\frac{d^{2}x^{k}}{d\sigma^{2}}&+&\frac{1}{c^{2}}\frac{\partial\Phi}{\partial
x^{j}}\left(c\frac{dt}{d\sigma}\right)^{2}+
\frac{1}{c^{2}}\frac{\partial\Phi}{\partial x^{k}}\delta_{ij}\frac{dx^{i}}{d\sigma}\frac{dx^{j}}{d\sigma}\label{eq:dduexk}\\
& &-\frac{2}{c^{2}}\frac{\partial\Phi}{\partial
x^{l}}\frac{dx^{l}}{d\sigma}\frac{dx^{k}}{d\sigma}+\frac{4}{c^{3}}\left(\frac{\partial
V^{k}}{\partial x^{j}}-\delta_{jm}\frac{\partial V^{m}}{\partial
x^{k}}\right)c\frac{dt}{d\sigma}\frac{dx^{i}}{d\sigma}=0\;,\nonumber
\end{eqnarray}
for the spatial components.

In the case of a null-geodesic, it is  $ds^{2}=d\sigma^{2}=0$. Eq.
(\ref{eq:ds_DUE}) gives, up to the order $c^{-3}$,

\begin{equation}
cdt=\frac{4V^{l}}{c^{3}}dx^{l}+\left(1-\frac{2\Phi}{c^{2}}\right)dl_{euclid}\;,\label{eq:c_dt}\end{equation}

where $dl_{euclid}^{2}=\delta_{ij}dx^{i}dx^{j}$ is the Euclidean
length interval. Squaring Eq.(\ref{eq:c_dt}) and keeping terms up
to order $c^{-3}$, one finds

\begin{equation}
c^{2}dt^{2}=\left(1-\frac{4\Phi}{c^{2}}
\right)dl_{euclid}^{2}+\frac{8V^{l}}{c^{3}}dx^{l}dl_{euclid}\;.\label{eq:cdue_dtdue}\end{equation}

Inserting Eq.(\ref{eq:cdue_dtdue}) into Eq.(\ref{eq:dduexk}), one
gets, for the spatial components,

\begin{equation}
\frac{d^{2}x^{k}}{d\sigma^{2}}+\frac{2}{c^{2}}\frac{\partial\Phi}{\partial
x^{k}}\left(\frac{dl_{euclid}}{d\sigma}\right)^{2}-\frac{2}{c^{2}}\frac{\partial\Phi}{\partial
x^{l}}\frac{dx^{l}}{d\sigma}\frac{dx^{k}}{d\sigma}+\frac{4}{c^{3}}\left(\frac{\partial
V^{k}}{\partial x^{j}}-\delta_{jm}\frac{\partial V^{m}}{\partial
x^{k}}\right)\frac{dl_{euclid}}{d\sigma}\frac{dx^{j}}{d\sigma}=0\;.\label{eq:ddue_dsigma}\end{equation}

 Such an equation can be seen as a differential equation for
$dx^k/d\sigma$ which is the tangent 3-vector to the trajectory. On
the other hand, Eq.(\ref{eq:ddue_dsigma}) can be expressed in
terms of $l_{euclid}$ considered as a parameter. In fact, for null
geodesics and taking into account the lowest order in $v/c$,
$d\sigma$ is proportional to $dl_{euclid}$. From Eq.(\ref{time})
multiplied for ${\displaystyle \left(1+\frac{2}{c^2}\Phi\right)}$,
we have
\begin{equation}
\frac{d}{d\sigma}\left(\frac{dt}{d\sigma}+\frac{2}{c^2}\Phi\frac{dt}{d\sigma}-
\frac{4}{c^4}\delta_{im}V^m\frac{dx^i}{d\sigma}\right)=0\,,
\end{equation}
and then
\begin{equation}
\frac{dt}{d\sigma}\left(1+\frac{2}{c^2}\Phi\right)-
\frac{4}{c^4}\delta_{im}V^m\frac{dx^i}{d\sigma}=1\,,\label{constant}
\end{equation}
where, as standard, we have defined the affine parameter so that
the integration constant is equal to 1 \cite{weinberg}.
Substituting Eq.(\ref{eq:c_dt}) into Eq.(\ref{constant}), at
lowest order in $v/c$, we find
\begin{equation} \frac{dl_{euclid}}{c d\sigma}=1\,.\end{equation}
In the weak field regime, the spatial 3-vector, tangent to a given
trajectory, can be expressed as
\begin{equation} \frac{dx^k}{ d\sigma}=\frac{cdx^k}{dl_{euclid}}\,.\end{equation}
By defining
\begin{equation} e^k=\frac{dx^k}{dl_{euclid}}\,,\end{equation}
Eq.(\ref{eq:ddue_dsigma}) becomes
\begin{equation}
\frac{de^{k}}{dl_{euclid}}+\frac{2}{c^{2}}\frac{\partial\Phi}{\partial
x^{k}}-\frac{2}{c^{2}}\frac{\partial\Phi}{\partial
x^{l}}e^{l}e^{k}+\frac{4}{c^{3}}\left(\frac{\partial
V^{k}}{\partial x^{j}}-\delta_{jm}\frac{\partial V^{m}}{\partial
x^{k}}\right)e^{j}=0\;,\label{eq:e_dsigma}\end{equation} which can
be expressed in a vector form as
\begin{equation}
\frac{d\mathbf{e}}{dl_{euclid}}=-\frac{2}{c^2}\left[\nabla\Phi-\mathbf{e}(\mathbf{e}\cdot\nabla\Phi)\right]+\frac{4}{c^3}
\left[\mathbf{e}\wedge(\nabla\wedge\mathbf{V})\right]\label{vector}\,.
\end{equation}
The gravitomagnetic term is the second one in Eq.(\ref{vector})
and it is usually discarded since considered not relevant. This is
not true if $v/c$ is quite large as in the cases of tight binary
systems or point masses approaching to black holes.

Our task is now to  achieve explicitly the trajectories, in
particular the orbits, corrected by such  effects.

\section{Theory of orbits}
Orbits with gravitomagnetic effects can be obtained starting from
the classical Newtonian theory and then correcting it by
successive relativistic terms. Here we give, for the sake of
completeness,  a quick review of classical and relativistic theory
of orbits showing how gravitomagnetic effects are the  further
corrections to be taken into account. A detailed discussion of
classical and relativistic theory of orbits can be found in
\cite{roy,brumberg}.

\subsection{The Newtonian theory}
The motion of a test particle  in a spherically symmetric
Newtonian gravitational field,   can be achieved starting from a
variational principle where the Lagrangian is

\begin{equation}
\mathcal{L}=\frac{1}{2}v^{2}+\frac{GM}{r}\label{eq:Lnewton}\end{equation}
where the particle mass has been assumed unitary. The velocity, in
spherical coordinates, is

\begin{equation}
v^{2}=\dot{r}^{2}+r^{2}\dot{\theta}^{2}+r^{2}\sin^{2}\theta\dot{\varphi}^{2}\,.\label{eq:v2}\end{equation}
Here the dot denotes the ordinary derivatives with  respect to the
time. The Euler-Lagrange equations are easily derived. For
$\theta$- component, we have

\begin{equation}
\frac{d}{dt}\left(r^{2}\dot{\theta}\right)=r^{2}\sin\theta\cos\theta\dot{\varphi}^{2}\,,\label{eq:ddt}\end{equation}
where an  obvious solution is $\theta=\pi/2$; in fact the motion
is plane and the variable $\theta$ cannot be  taken in
consideration any more. The equation

\begin{equation}
\frac{d}{dt}\left(r^{2}\dot{\varphi}\right)=0\,,\end{equation}
gives

\begin{equation}
r^{2}\dot{\varphi}=const=H\,,\label{eq:H}\end{equation} which is
nothing else but the conservation  of the angular momentum.
Finally,  we have

\begin{equation}
\ddot{r}=r\dot{\varphi}^{2}-\frac{GM}{r^{2}}\,.\label{eq:rdotdot1}\end{equation}
It is convenient to introduce the new variable

\begin{equation}
u(\varphi)=\frac{1}{r}\label{eq:u}\end{equation}
 Being
\begin{equation}
u'=\frac{du}{d\varphi},\label{eq:uprimo}\end{equation} and using
Eq.(\ref{eq:u}) and Eq.(\ref{eq:H}), it results

\begin{equation}
\dot{r}=
-\frac{1}{u}\frac{du}{dt}=-r^{2}\frac{du}{d\varphi}\frac{d\varphi}{dt}=
-r^{2}\dot{\varphi}u'=-Hu'.\label{eq:rdot}\end{equation} From this
equation, one gets

\begin{equation}
\ddot{r}=-H\frac{d}{dt}\left(\frac{du}{d\varphi}\right)=-H\frac{d\varphi}{dt}\frac{d}{d\varphi}\left(\frac{du}{d\varphi}\right)=-H\dot{\varphi}u''=-\frac{H^{2}}{r^{2}}u''=-H^{2}u^{2}u''\label{eq:rdotdot}\end{equation}
and then Eq.(\ref{eq:rdotdot1}) is

\begin{equation}
u''+u=\frac{GM}{H^{2}}\label{eq:usec}\end{equation} where the
trivial solution $u=0$ ($r=\infty$) is discarded. The solution of
Eq.(\ref{eq:usec}) is

\begin{equation}
u=\frac{GM}{H^{2}}+B\cos(\varphi-\varphi_{0}),\label{eq:solu}\end{equation}
and then,  imposing $\varphi_{0}=0$, one gets the orbits in polar
coordinates

\begin{equation}
r(\varphi)=\frac{k}{1+e\cos\varphi}\,.\label{eq:solu1}\end{equation}

Here ${\displaystyle k=\frac{GM}{H^{2}}}$ and $e$ is the
ellipticity whose value can give elliptic, hyperbolic and
parabolic orbits \cite{landau1}. Summarizing the solution for
$\theta$ gives the planar motion, the solution for $\varphi$ gives
 the angular momentum conservation, while the solution for $r$
gives  the orbits.

\subsection{The relativistic theory}

The relativistic case  can be seen as a correction to the
Newtonian theory of orbits. As before, we can start from a
Lagrangian which can be deduced from the Schwarzschild line
element, that is

\begin{equation}
\mathcal{L}=e^{\nu}\left(\dot{x}^{0}\right)^{2}-
e^{\lambda}\left(\dot{r}\right)^{2}-r^{2}\left(\dot{\theta}^{2}+\sin^{2}\theta\dot{\varphi}^{2}\right).\label{eq:Lrel}\end{equation}

The Euler-Lagrange equation  for $\theta$ is

\begin{equation}
\frac{d}{ds}\left(r^{2}\dot{\theta}\right)=r^{2}\sin\theta\cos\theta\dot{\varphi}^{2}.\label{eq:dds}\end{equation}
In  analogy with Eq. (\ref{eq:ddt}) (the two equations differ for
$ds$ in place of $dt$), the solution of this equation is
$\theta=\pi/2$; again, as in the classical case, the motion  is
plane and $\theta$ disappears as dynamical variable. The equations
for $x^{0}=ct$ and $x^{3}=\varphi$ admit two first integrals of
motion since the Lagrangian does not depend on $x^{0}$ and on
$x^{3}$ but only on their derivatives. We have

\begin{equation}
\left(1-\frac{R_{s}}{r}\right)\dot{x}_{0}=l,\qquad
r^{2}\dot{\varphi}=h,\label{eq:lh}\end{equation} corresponding to
the first integrals of energy and   angular momentum. $R_{s}$ is
the Schwarzschild radius. For $x^{1}=r$ we can use the definition
$\mathcal{L}=g^{\mu\nu}\dot{x}_{\mu}\dot{x}_{\nu}=1$ instead of
the corresponding second order equation. Being ${\displaystyle
e^{\nu}=e^{-\lambda}=\left(1-\frac{R_{s}}{r}\right)}$, we have

\begin{equation}
\mathcal{L}=\left(1-\frac{R_{s}}{r}\right)\left(\dot{x}^{0}\right)^{2}-
\frac{\left(\dot{r}\right)^{2}}{\left(1-\frac{R_{s}}{r}\right)}-
r^{2}\left(\dot{\theta}^{2}+\sin^{2}\theta\dot{\varphi}^{2}\right)=1.\label{eq:Lrelnew}\end{equation}
Replacing Eq. (\ref{eq:lh}) and considering $\theta=\pi/2$, we
have

\begin{equation}
l^{2}-\dot{r}^{2}-\frac{h^{2}}{r^{2}}\left(1-\frac{R_{s}}{r}\right)=\left(1-\frac{R_{s}}{r}\right)\,.
\label{eq:l2}\end{equation}

As in the Newtonian case, using the  variable given by
Eq.(\ref{eq:u}) and using the second of Eqs.(\ref{eq:lh}), it is

\begin{equation}
\dot{r}=-hu'\label{eq:rdotrel}\,.\end{equation}

Inserting Eq.(\ref{eq:rdotrel}) and Eq.(\ref{eq:u}) in
Eq.(\ref{eq:l2}), we get

\begin{equation}
l^{2}-h^{2}u'-h^{2}u^{2}\left(1-R_{s}u\right)=\left(1-R_{s}u\right).\label{eq:l2new}\end{equation}

This  equation gives, by a quadrature, the solution $u=u(\varphi)$
with the periastron precession but, in order to compare the result
with the Newtonian case, we can derive Eq.(\ref{eq:l2new})
considering that $\ddot{r}=-hu^{2}u''$. One obtains

\begin{equation}
u''+u=\frac{R_{S}}{2h^{2}}+\frac{3}{2}R_{S}u^{2}\label{eq:usecrel}\end{equation}
This equation can be easily  compared with the corresponding
Newtonian case (\ref{eq:usec}) since

\begin{equation}
h\simeq
r^{2}\frac{1}{c}\dot{\varphi}=\frac{H}{c}.\label{eq:h}\end{equation}
Being ${\displaystyle R_{S}=\frac{2GM}{c^{2}}}$, it follows that

\begin{equation}
\frac{R_{S}}{2h^{2}}\simeq\left(\frac{GM}{c^{2}}\right)\left(\frac{c^{2}}{H^{2}}\right)=
\left(\frac{GM}{c^{2}}\right)\,.\label{eq:Rssu2h}\end{equation}
This means that  the relativistic correction  to the test particle
motion  is due to the second member of (\ref{eq:usecrel}). Such a
term is small is small if compared to the other. In fact, using
(\ref{eq:h}) we have

\begin{equation}
\frac{\frac{3}{2}R_{S}u^{2}}{\frac{R_{S}}{2h^{2}}}=3h^{2}u^{2}\simeq\frac{3H^{2}}{r^{2}c^{2}}=
3\left(\frac{v}{c}\right)^{2}\,,\end{equation} so we can use a
perturbation approach to deal with it. As said, such a
relativistic correction is responsible for the perihelion
precession. However, in strong field and high relative velocity
regime, such term has relevant effects.

\subsection{Relativistic corrections due to gravitomagnetic effects}
Starting from the above considerations, we can see how
gravitomagnetic corrections affect the problem or orbits.
Essentially, they act as a further $v/c$ correction leading to
take into account terms up to $c^{-3}$, as shown in Sec.II.

Let us start from the line element (\ref{eq:ds_DUE}) which can be
written in spherical coordinates. Here we assume the motion of
point-like bodies and then we can work in the simplified
hypothesis ${\displaystyle \Phi=-\frac{GM}{r}}$ and $V^{l}=\Phi
v^{l}$. It is

\begin{eqnarray*}
ds^{2} & = &
\left(1+\frac{2\Phi}{c^{2}}\right)cdt^{2}-\left(1-\frac{2\Phi}{c^{2}}\right)\left[dr^{2}+r^{2}d\theta^{2}+
r^{2}\sin^{2}\theta d\varphi^{2}\right]
 -\frac{8\Phi}{c^{3}}cdt \left\{\left[\cos\theta+\sin\theta\left(\cos\varphi+\sin\varphi\right)\right]dr\right.\\
 &  & \left. +\left[\cos\theta\left(\cos\varphi+\sin\varphi\right)-\sin\theta\right]rd\theta
 +\left[\sin\theta\left(\cos\varphi-\sin\varphi\right)\right]rd\varphi\right\}\,.
 \label{totalmetricnostrpola}\end{eqnarray*}
As in the Newtonian and relativistic cases, from the line element
(\ref{totalmetricnostrpola}), we can construct the Lagrangian

 \begin{eqnarray}
\mathcal{L} & = &
\left(1+\frac{2\Phi}{c^{2}}\right)\dot{t}^{2}-\left(1-\frac{2\Phi}{c^{2}}\right)\left[\dot{r}^{2}+
r^{2}\dot{\theta}^{2}+r^{2}\sin^{2}\theta \dot{\varphi}^{2}\right]
  -\frac{8\Phi\dot{t}}{c^{3}} \left\{ \left[\cos\theta+\sin\theta\left(\cos\varphi+\sin\varphi\right)
  \right]\dot{r}\right.\nonumber\\
 &  & \left.+\left[\cos\theta\left(\cos\varphi+\sin\varphi\right)-\sin\theta\right]r\dot{\theta}
  +\left[\sin\theta\left(\cos\varphi-\sin\varphi\right)\right]r\dot{\varphi}\right\}
 \label{Lagrangiannostra}\,.\end{eqnarray}
Using the relations (\ref{eq:lh}) and being, as above,
$\mathcal{L}=1$, one can multiply both  members for
${\displaystyle \left(1+\frac{2\Phi}{c^{2}}\right)}$. In the
planar motion condition $\theta=\pi/2$ , we  obtain

\begin{equation}
l^{2}-\left(1+\frac{2\Phi}{c^{2}}\right)\left(1-\frac{2\Phi}{c^{2}}\right)\left(\dot{r}^2+\frac{h^{2}}{r^{2}}\right)
-\frac{8\Phi
l}{c^{3}}\left[\left(\cos\varphi+\sin\varphi\right)\dot{r}-\left(\cos\varphi-\sin\varphi\right)\dot{\varphi}\right]
 = \left(1+\frac{2\Phi}{c^{2}}\right)\,,\end{equation} and then,
being ${\displaystyle \frac{2\Phi}{c^2}=-\frac{R_{s}}{r}}$ and
${\displaystyle u=\frac{1}{r}}$ it is

\begin{equation}
l^{2}-h^{2}\left(1-R_{s}^{2}u^{2}\right)\left(u'^{2}+u^{2}\right)+
\frac{4R_{S}ul}{c}\left[\left(\cos\varphi+\sin\varphi\right)u'+\left(\cos\varphi-\sin\varphi\right)u^{2}\right]
 =  \left(1-R_{S}u\right)\label{eq:L4}\,.\end{equation}
By deriving such an equation, it is easy to show that, if the
relativistic and gravitomagnetic terms are discarded, the
Newtonian theory is recovered, being
\begin{equation}
u''+u=\frac{R_s}{2h^2}\,.\end{equation} This result probes the
self-consistency of the problem. However, it is nothing else but a
particular case since we have assumed the planar motion. This
planarity condition does not hold in general  if  gravitomagnetic
corrections are taken into account.

\section{Orbits with gravitomagnetic effects}

From the above Lagrangian (\ref{Lagrangiannostra}), it is
straightforward to derive the equations of motion

  \begin{eqnarray}
 \ddot{r}&=&\frac{1}{c r \left(r c^2+2 G M\right)}
 \Big[c \left(r c^2+G M\right) \left(\dot{\theta}^2+\sin ^2\theta \dot{\phi}^2\right) r^2\\
 \nonumber
 && -4 G M \dot{t} \left((\cos\theta(\cos\phi+\sin\phi)-\sin \theta)\dot{\theta}+
 \sin\theta (\cos\phi-\sin \phi)\dot{\phi} \right) r+c G M \dot{r}^2-c G M
 \dot{t}^2\Big]\,,
  \end{eqnarray}

 \begin{eqnarray}
\ddot{\phi}=-\frac{2 \left(c \cot \theta \left(r c^2+2 G M\right) \dot{\theta}\dot{ \phi} r^2+\dot{r}
   \left(2 G M \csc \theta (\sin\phi-\cos\phi) \dot{t}+c r \left(r c^2+G
   M\right) \dot{\phi}\right)\right)}{r^2 \left(r c^3+2 G M
   c\right)}\,,
   \end{eqnarray}

\begin{eqnarray}
 \ddot{\theta}=\frac{c \cos\theta r^2 \left(r c^2+2 G M\right) \sin\theta \dot{\phi}^2+\dot{r}
   \left(4 G M (\cos\theta (\cos \phi+\sin\phi)-\sin\theta)\dot{t}-
   2 c r \left(r c^2+G M\right) \dot{\theta}\right)}{r^2 \left(r c^3+2 G M
   c\right)}\,,\end{eqnarray}
corresponding to the spatial components of the geodesic Eq.
(\ref{eq:ddue_dsigma}). Due to the numerical calculations which we
are going to perform below, we consider the explicit form of the
equations of motion. We have not considered the time component
$\ddot{t}$ since it is not necessary for the discussion of orbital
motion.

As remarked above, from $\mathcal{L}=1$  the first integral
$\dot{r}$ is achieved. It is:
 \begin{eqnarray}
 \nonumber
 \dot{r}=\frac{1}{r^2 c^6+4 G^2 M^2 \left(-c^2+4 \sin 2 \theta (\cos\phi+\sin\phi)+4 \sin^2\theta \sin 2 \phi+4\right)}\pm \Bigg[ r\Big(64 G^4 M^4 r \\
 \nonumber
\left((2 \cos 2 \theta (\cos\phi+\sin\phi)+\sin 2 \theta \sin 2 \phi) \dot{\theta} +\left(2 \cos 2 \phi \sin ^2\theta+\sin 2 \theta (\cos\phi-\sin\phi)\right)\dot{\phi}\right)^2+\\
 \nonumber
  - \Big(r^2 c^6+4 G^2 M^2 \big(-c^2+4 \sin 2 \theta (\cos\phi+\sin\phi)+4 \sin ^2\theta \sin 2 \phi
  +4\big)\Big) \Big(r^3 \big(\dot{\theta}^2+\sin ^2\theta \dot{\phi}^2\big) c^6-4 G M c^4+\\
  \nonumber
  -r \Big(\big(\bold{E}^2-2) c^6+4 G^2 M^2 \Big(\big(c^2+4 \sin 2 \theta (\cos\phi+\sin\phi)
  -4 \cos ^2\theta \sin 2 \phi-4\Big)\dot{\theta}^2-8 \sin\theta (\cos\phi-\sin\phi)\\
  \nonumber
   \bigg(\cos\theta (\cos\phi+\sin \phi-\sin\theta) \dot{\phi} \dot{\theta}+\sin ^2\theta \big(c^2+4 \sin 2 \phi-4\big)\dot{\phi}^2)\Big)\bigg)\bigg)-8 G^2 M^2 r \Big((2 \cos 2 \theta (\cos\phi+\sin\phi)+\\
   \sin 2 \theta \sin 2 \phi)\dot{ \theta}+\big(2 \cos 2 \phi \sin ^2\theta+\sin 2 \theta
   (\cos\phi-\sin \phi)\big)\dot{\phi}\Big)\Bigg]^{\frac{1}{2}}\,,
 \end{eqnarray}
which is the natural constrain equation related to the energy. The
double sign comes out from the quadratic form of the Lagrangian.
For our purpose, the positive sign can be retained.

In the following calculations, we adopt geometrized  units. Our
aim is  to study how gravitomagnetic effects modify the orbital
shapes and what are the parameters determining the stability of
the problem. As we will see, the energy and the mass, essentially,
determine the stability. Beside the standard periastron precession
of General Relativity, a nutation effect is genuinely induced by
gravitomagnetism and stability greatly depends on it. A
fundamental issue for this study is to achieve the orbital phase
space portrait.

\section{Numerical results}

The  solution of the above system  of differential equations
presents some difficulties since the equations are  stiff and
their numerical solutions can diverge in several test  points.
Some numerical algorithms allow to change dynamically the meshing
in order to decrease the mesh size near the critical points.

For our purposes, we have found  solutions by using the so called
{\it Stiffness Switching Method} to provide an automatic tool of
switching between a non-stiff and a stiff solver coupled with a
more conventional explicit Runge-Kutta method for the non-stiff
part of our differential equations.

\begin{figure}[!ht]
\begin{tabular}{|c|}
\hline
\includegraphics[scale=0.8]{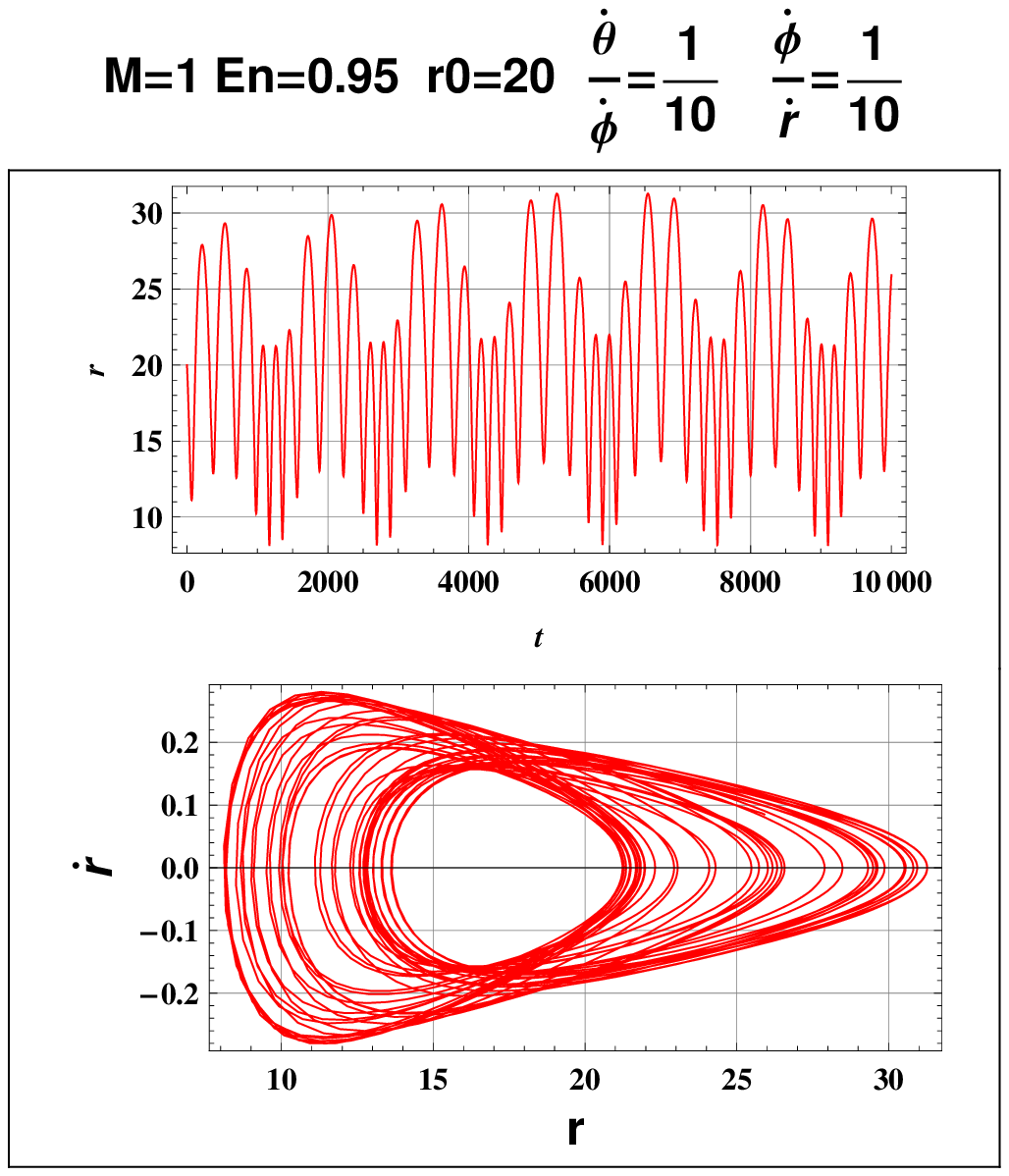} \tabularnewline
\hline
\includegraphics[scale=0.8]{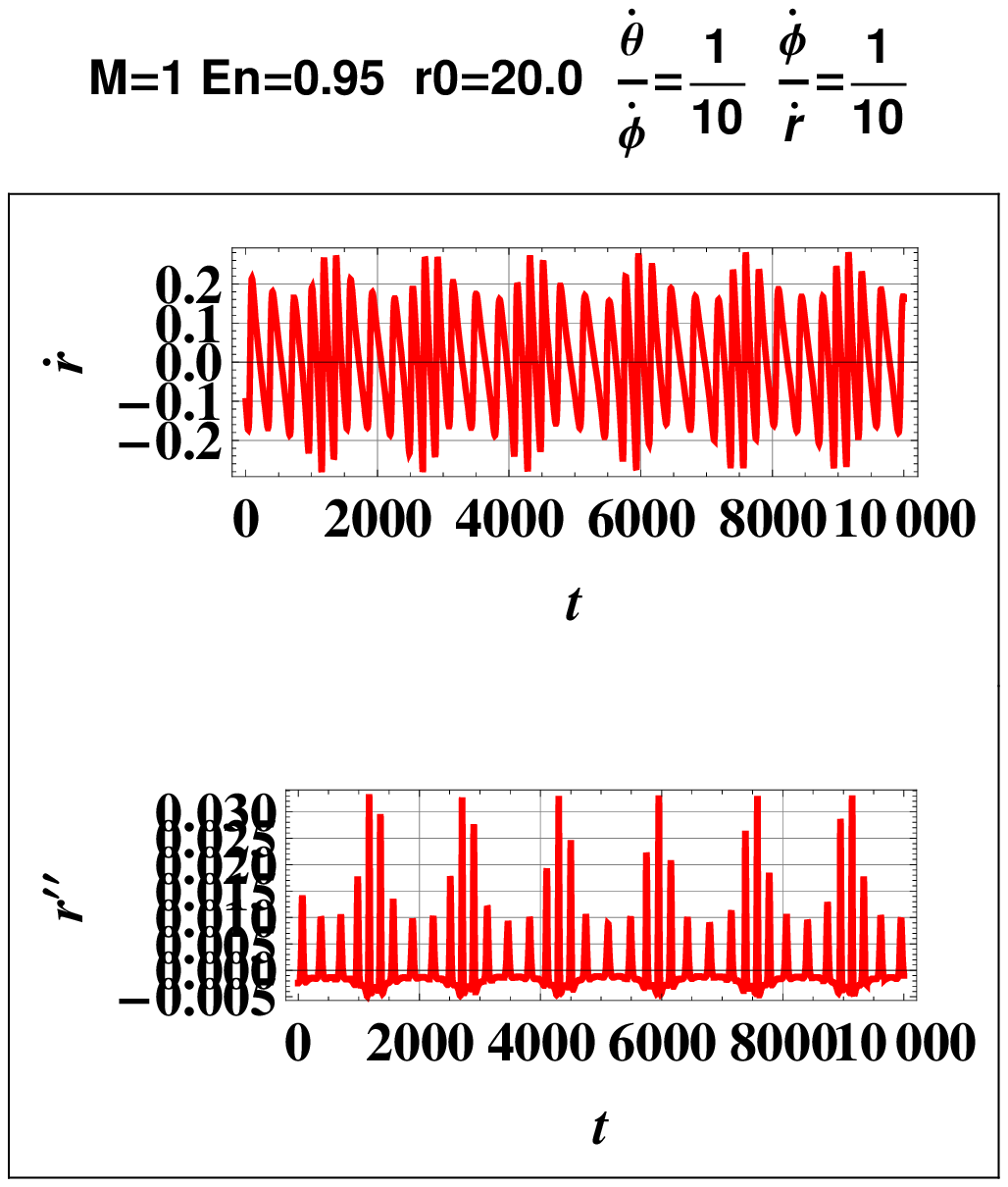} \tabularnewline
\hline
\end{tabular}
\caption {Plots of $\dot{r}(t)$  and $\ddot{r}(t)$  for a test
mass $M=1 M_{\bigodot}$, energy per mass unit $E_n=0.95$ and
initial values for the orbital radius $r_{0}=20$, given in terms
of Schwarzschild radius. The initial values of the angular
precession velocity $\dot{\phi}$ and the angular nutation velocity
$\dot{\theta}$ have been chosen according to the following
criterium: assuming a given value of the initial radial velocity
$\dot{r}$,  the initial values of the angular precession velocity
and of the angular nutation velocity are
$\dot{\phi}=-\frac{1}{10}\dot{r}$ and $\dot{\theta}=
-\frac{1}{100}\dot{r}= \frac{1}{10}\dot{\phi}$. The phase portrait
of $\dot{r}=f(r)$ is shown. The adopted time span is 10000 steps.}
\label{Fig:stiffness}
\end{figure}

\begin{figure}[!ht]
\begin{tabular}{|c|c|}
\hline
\includegraphics[height=0.2\textheight]{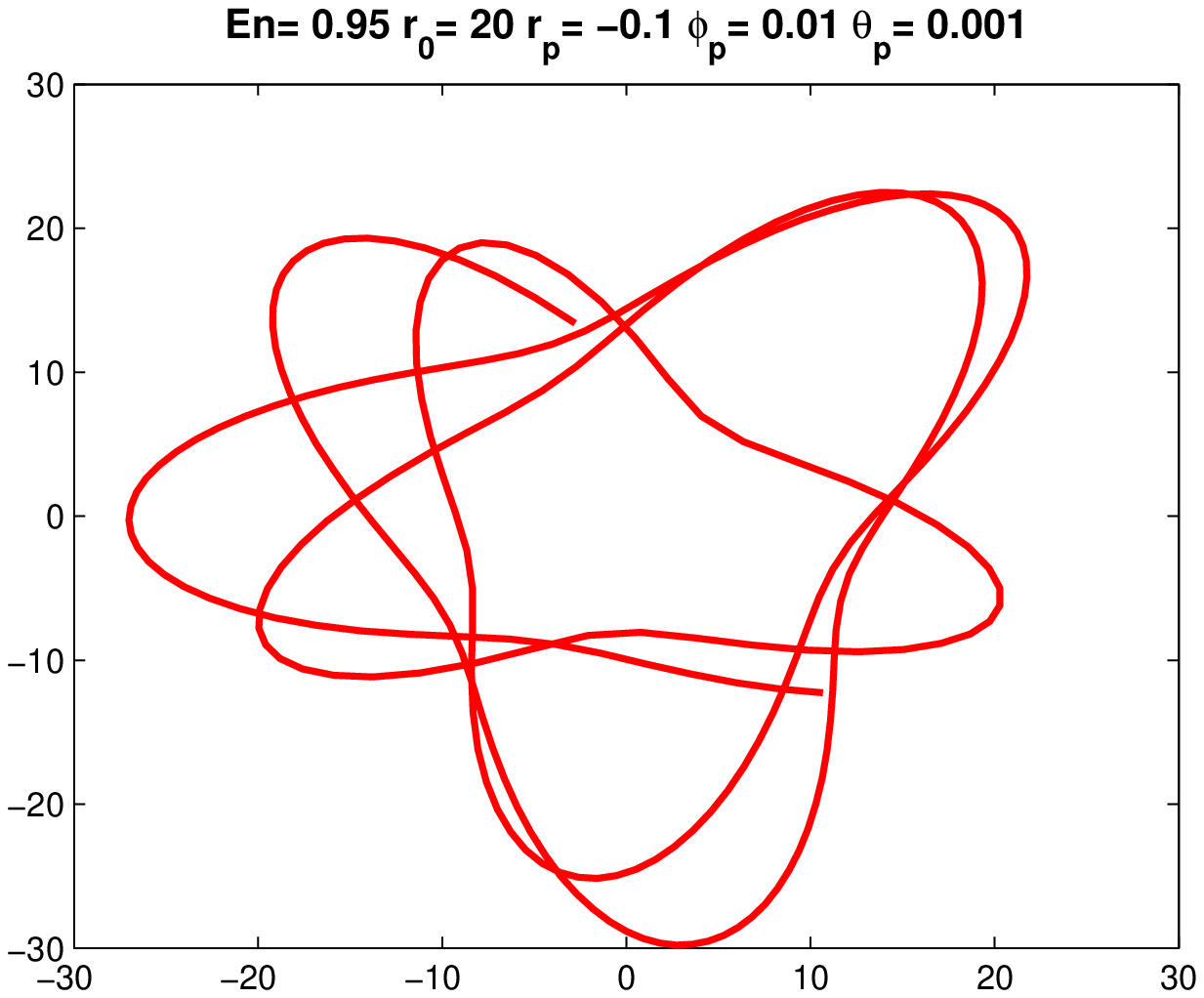}&
\includegraphics[height=0.2\textheight]{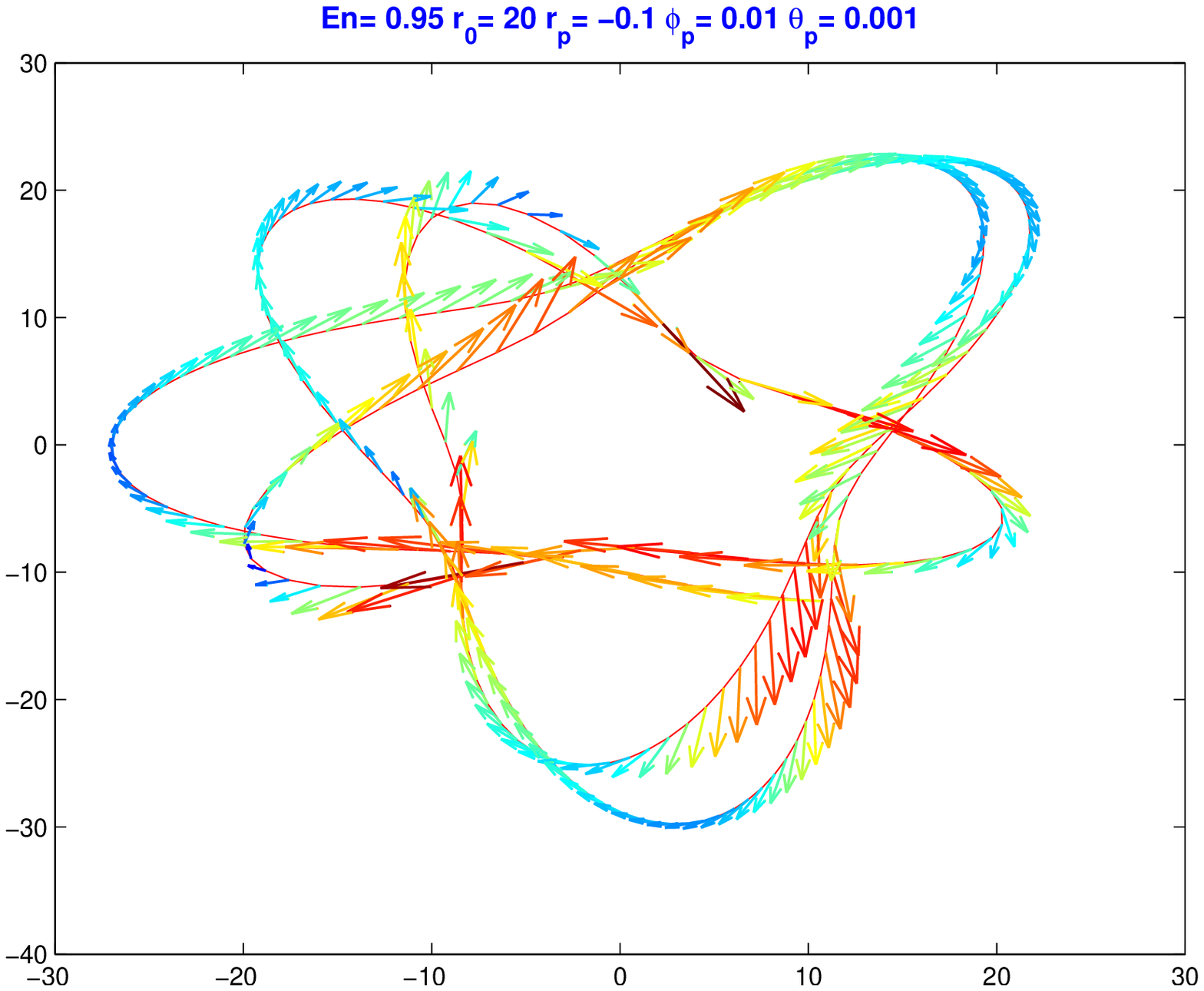}\tabularnewline
\hline
\end{tabular}
\caption {Plots of basic orbits (left) and  orbits with the
associated velocity field (right). The arrows indicate the
instantaneous velocities. The initial values are:
$M=1M_{\bigodot}$; $E_n=0.95$ in mass units;
 $r_0=20$ in Schwarzschild radii; $\dot\phi=-\frac{\dot
r}{10}$; $\dot\theta=\frac{\dot\phi}{10}$. } \label{fig:orbit}
\end{figure}

\begin{figure}[!ht]
\begin{tabular}{|c|c|}
\hline
\includegraphics[height=0.2\textheight]{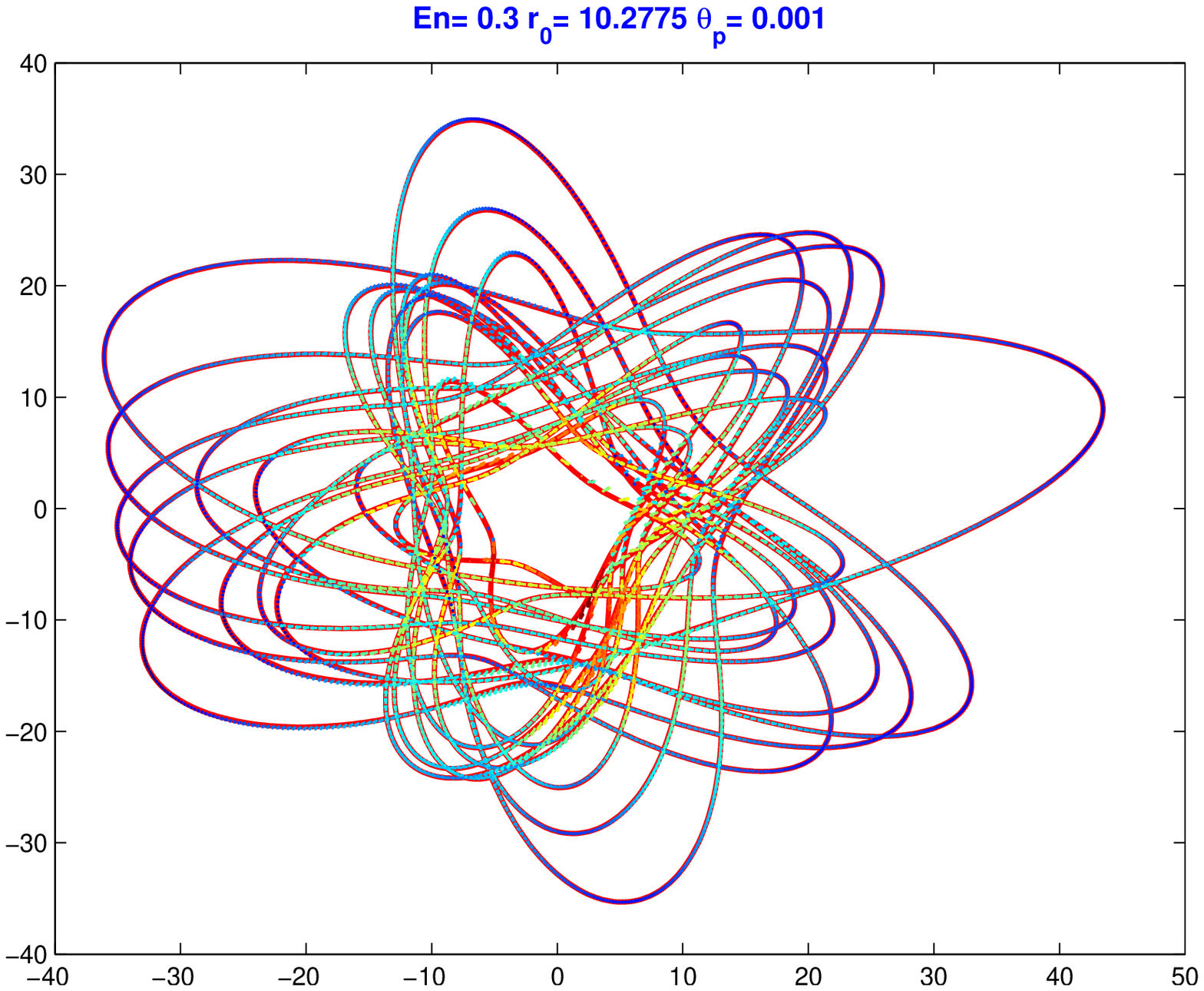}&
\includegraphics[height=0.2\textheight]{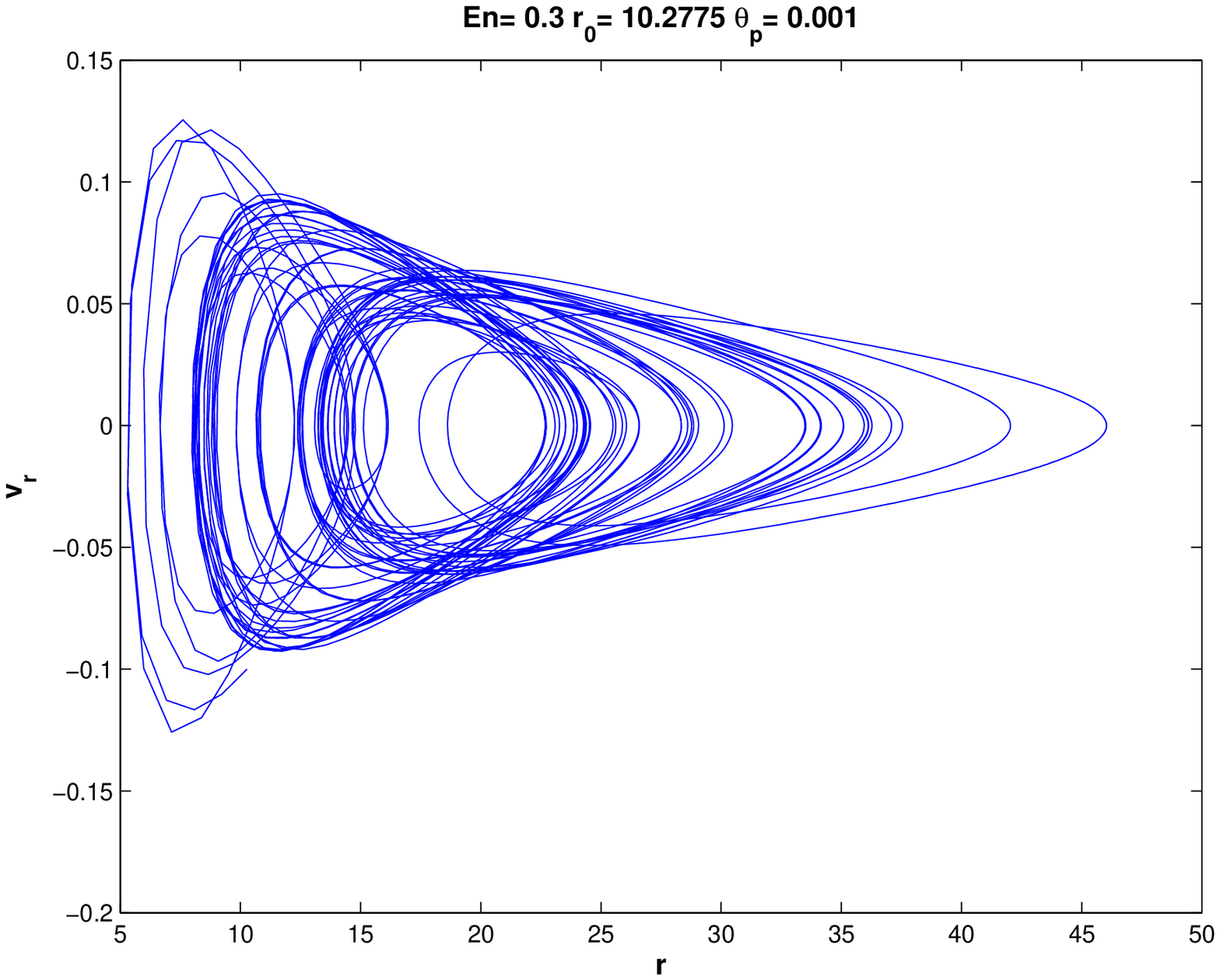}\tabularnewline
\hline
\includegraphics[height=0.2\textheight]{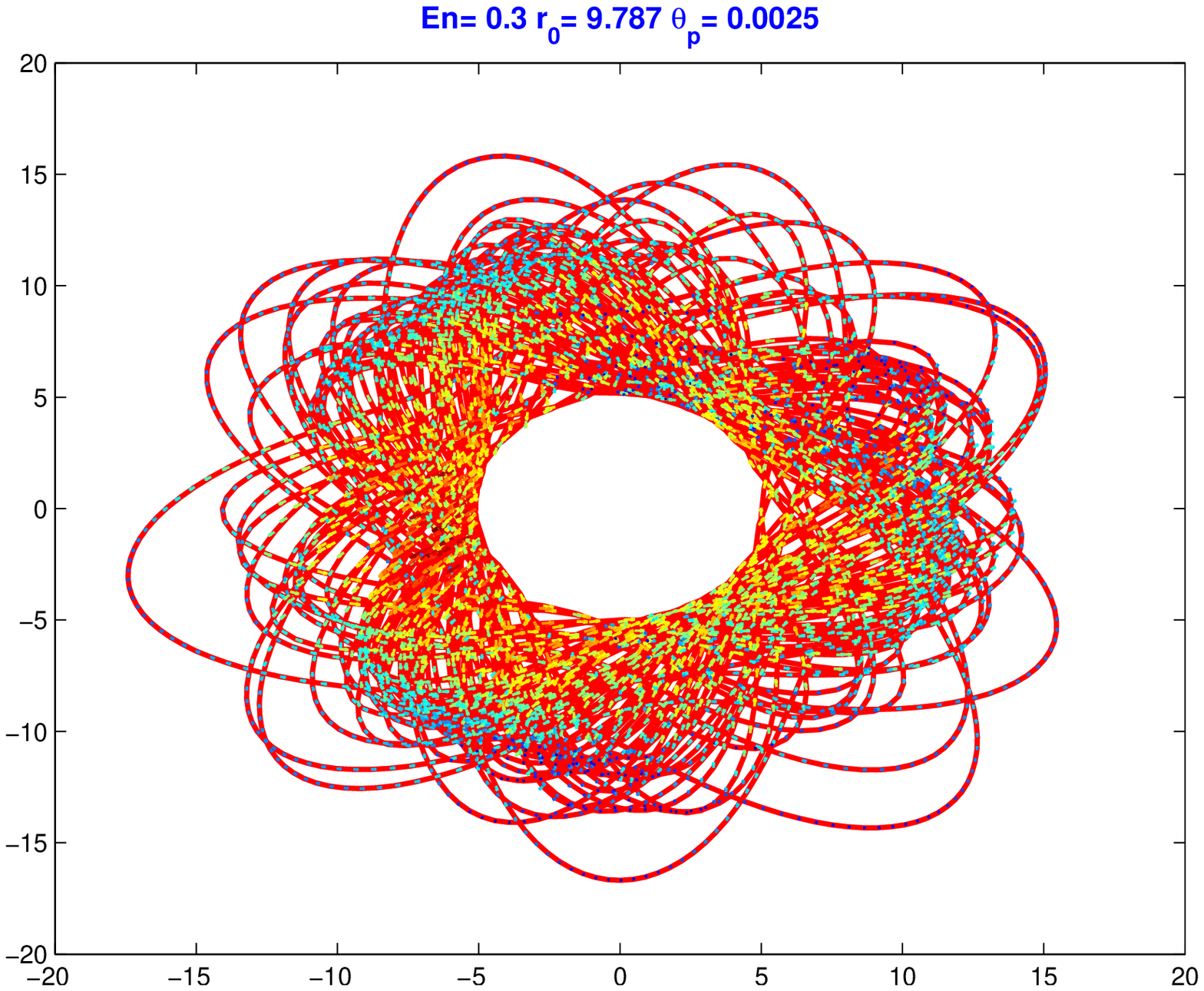}&
\includegraphics[height=0.2\textheight]{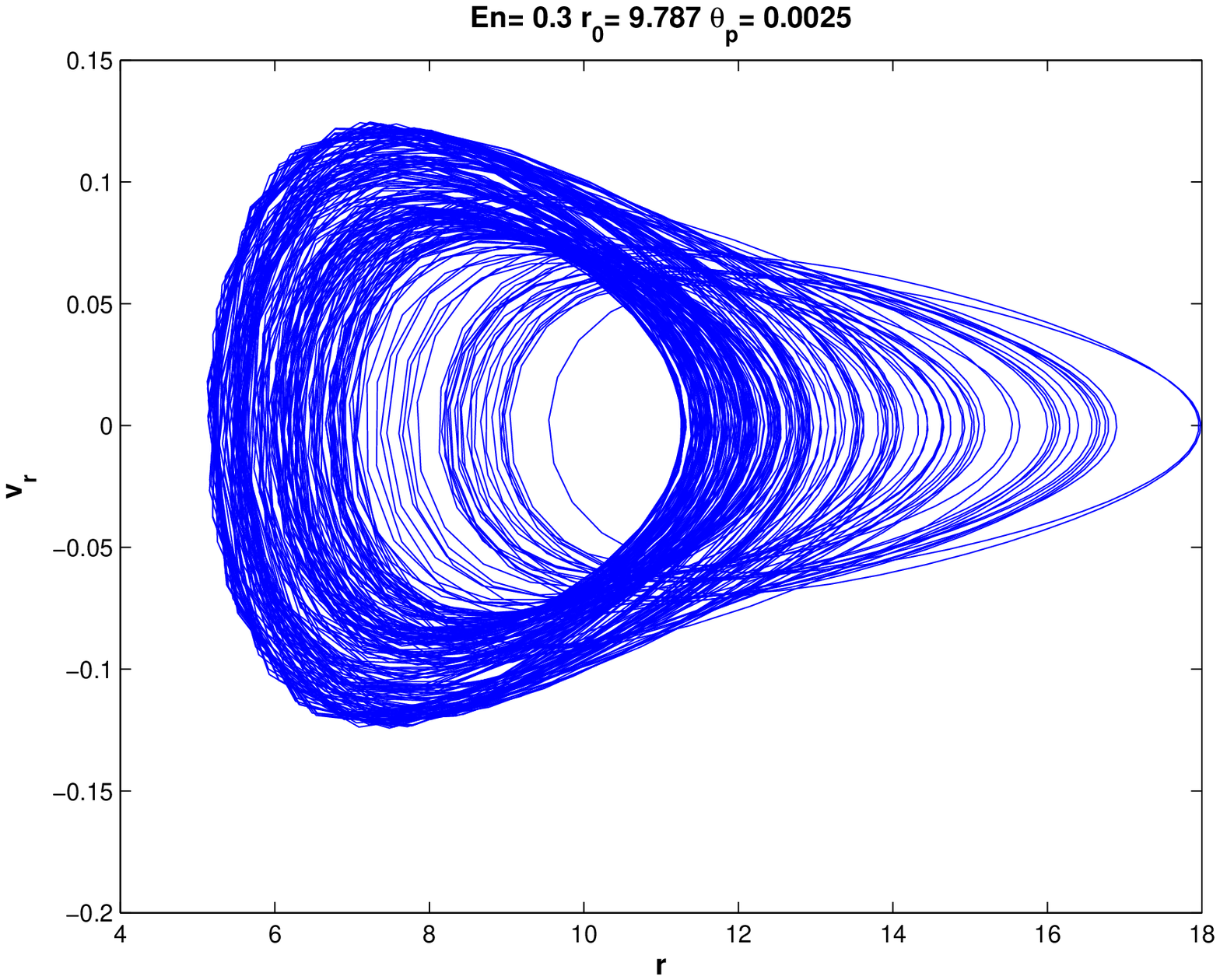}\tabularnewline
\hline
\includegraphics[height=0.2\textheight]{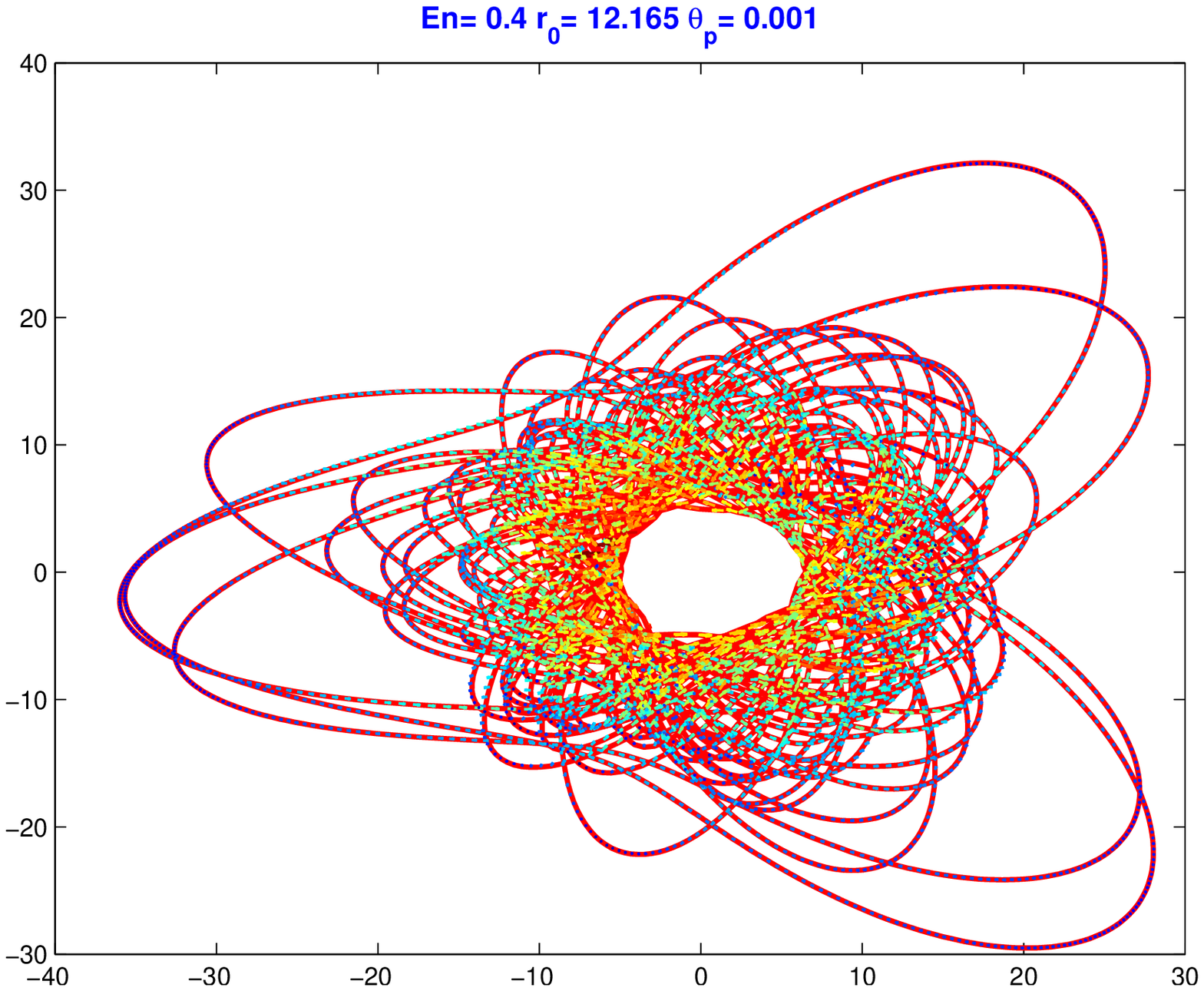}&
\includegraphics[height=0.2\textheight]{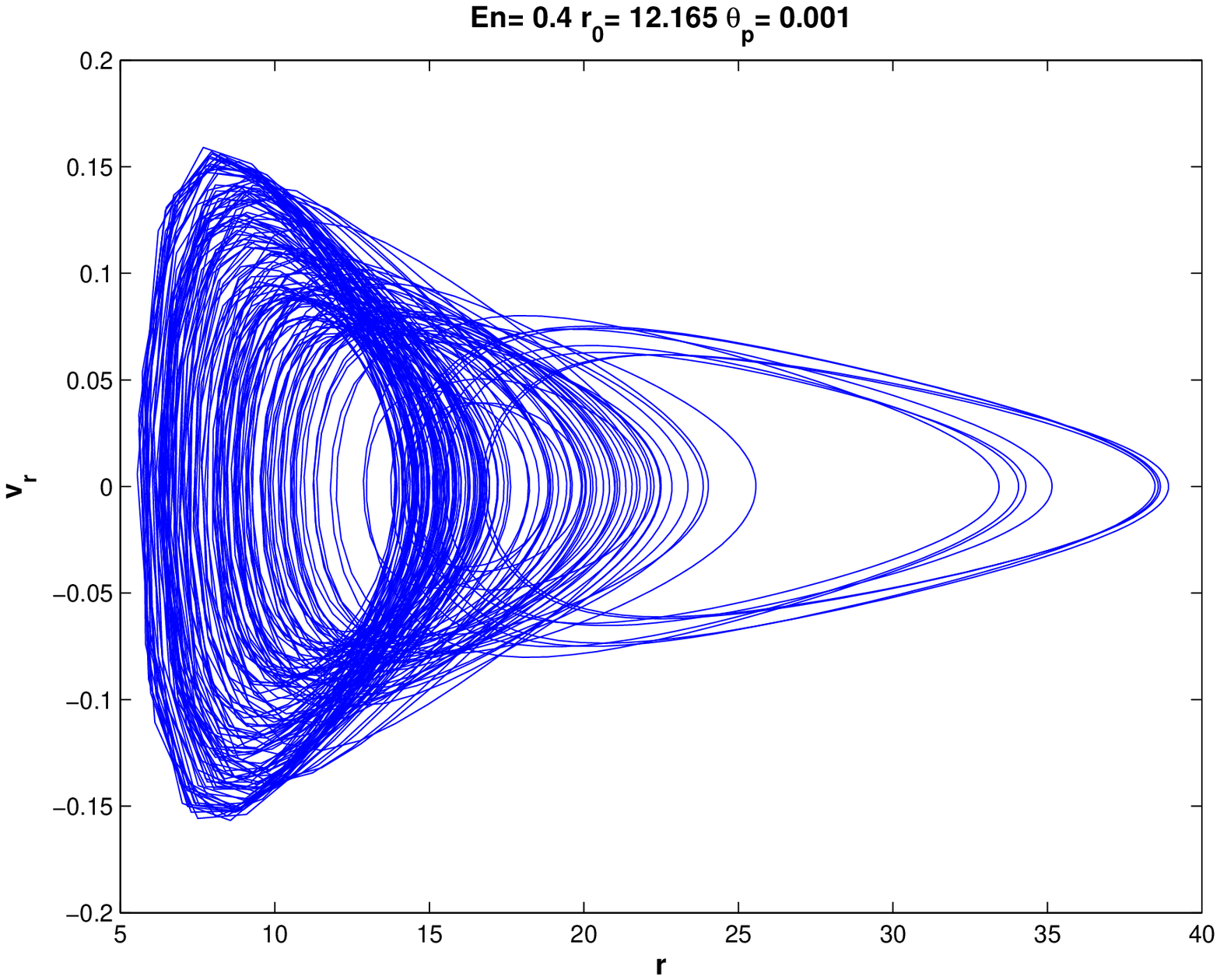}\tabularnewline
\hline
\includegraphics[height=0.2\textheight]{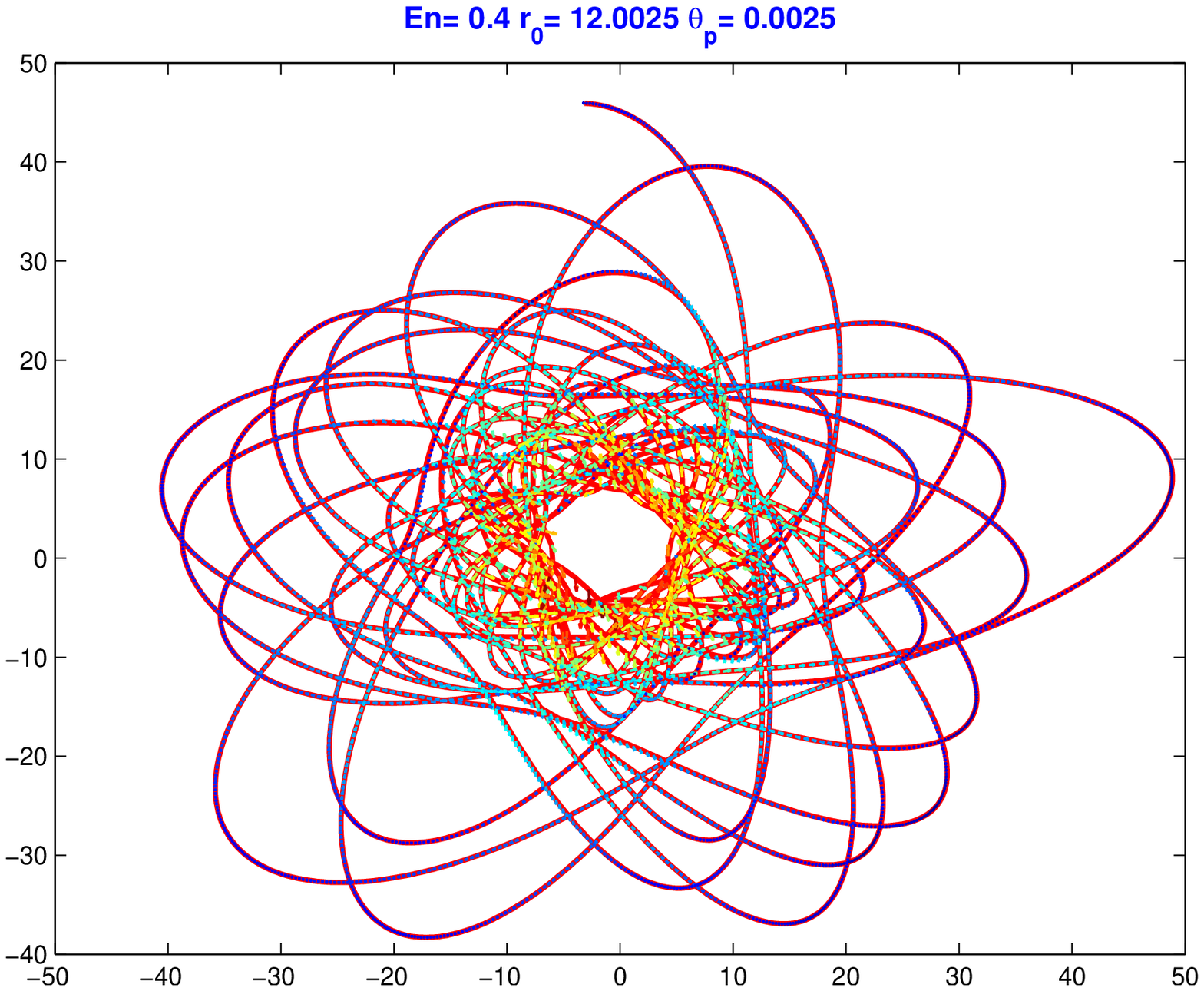}&
\includegraphics[height=0.2\textheight]{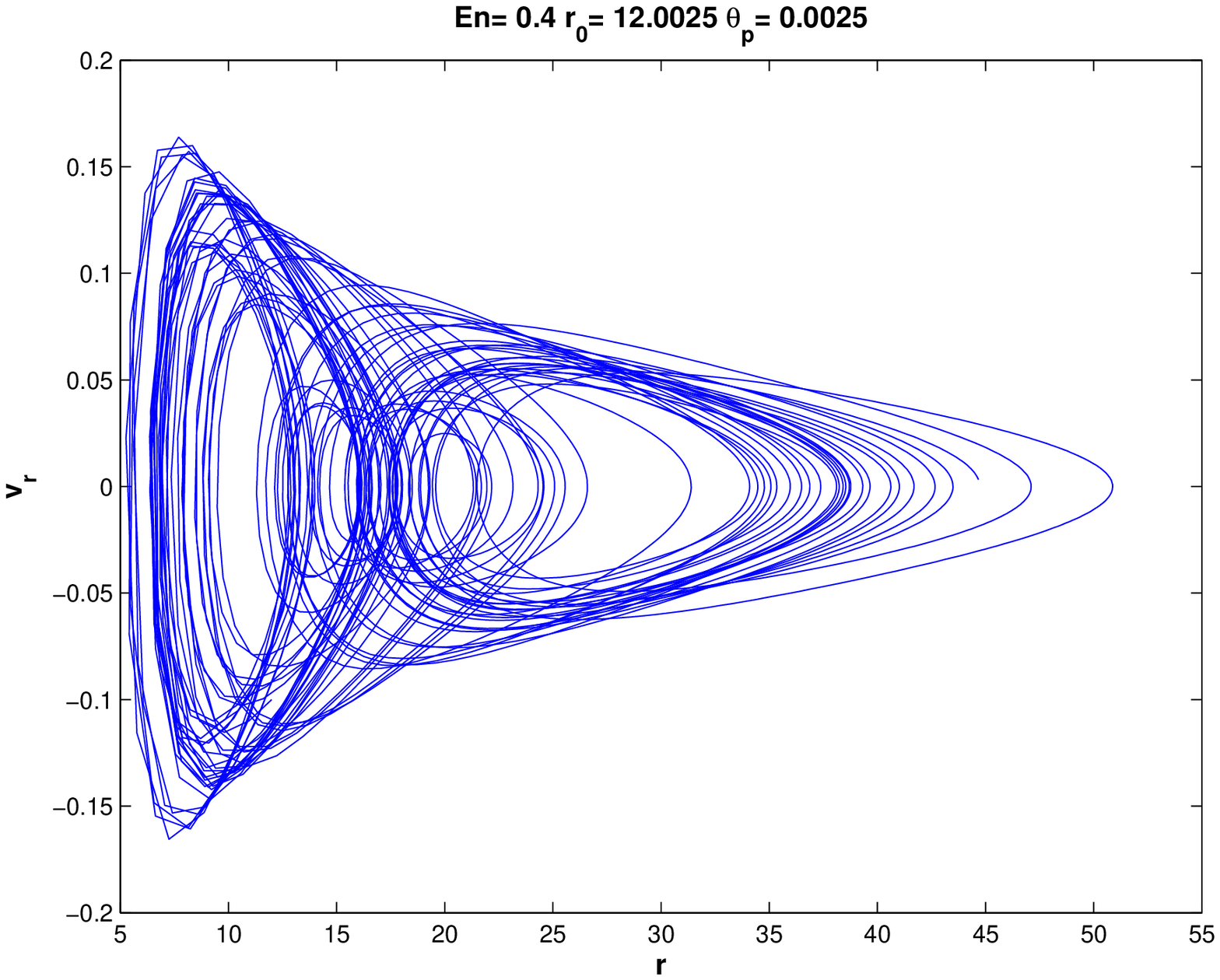}\tabularnewline
\hline
\end{tabular}
\caption {Plots of orbits with  various  energy values. For each
value of  energy, four plots are shown: the first on the left
column is the orbit, with the orbital velocity field in false
colors. The color scale goes from blue to red in increasing
velocity. The second on the left column is the orbit with a
different nutation angular velocity. On the right column, the
phase portraits $\dot r=\dot r(r(t))$ are shown. Energy varies
from $0.3$ to $0.4$, in mass units. The stability of the system is
highly sensitive either to very small variation of $r_{0}$ or to
variation on the initial conditions on both precession and
nutation angular velocities: it is sufficient a variation of few
percent on $r_{0}$ to induce system instability} \label{fig:1}
\end{figure}

\begin{figure}[!ht]
\begin{tabular}{|c|c|}
\hline
\includegraphics[scale=0.5]{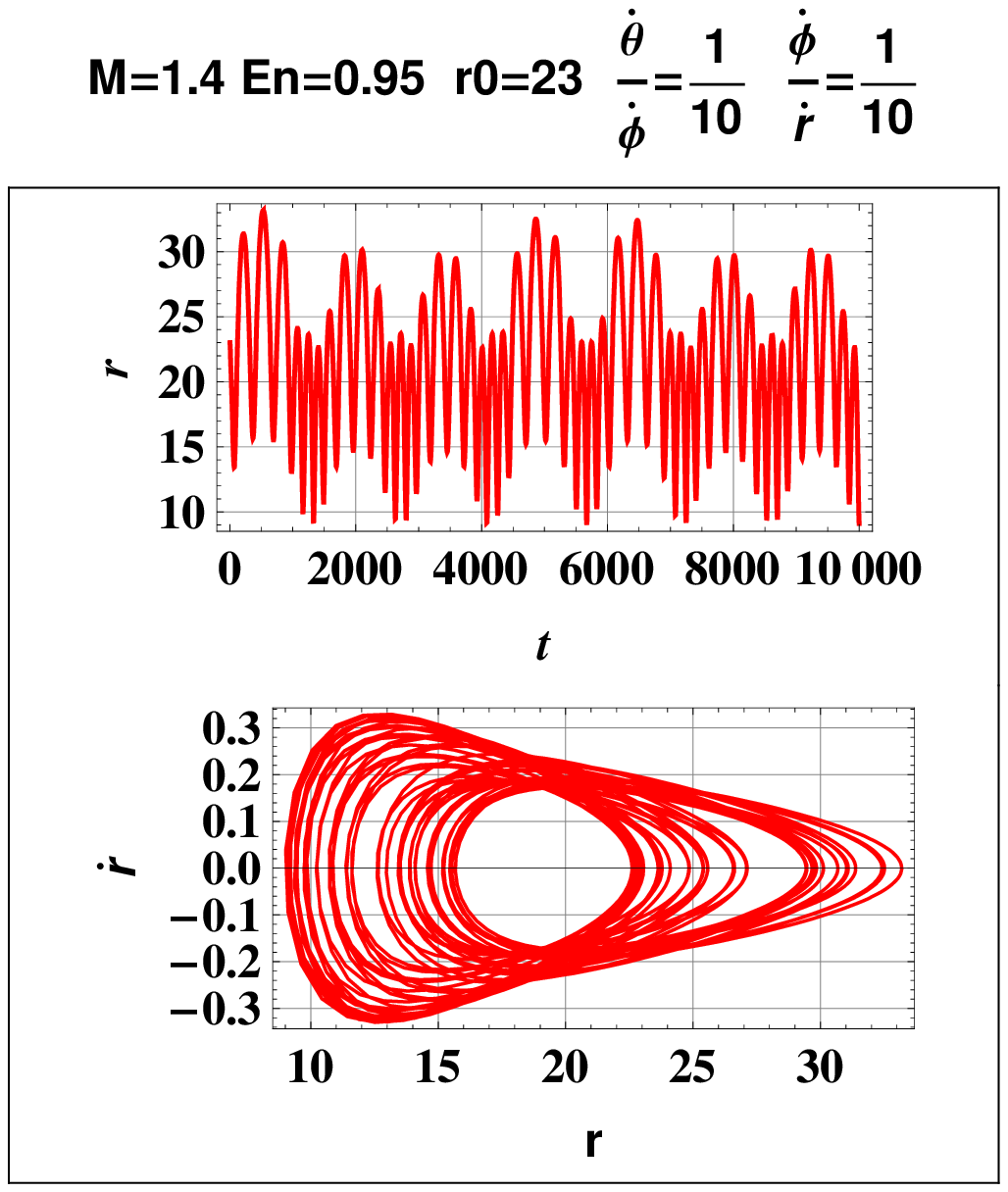}&
\includegraphics[scale=0.5]{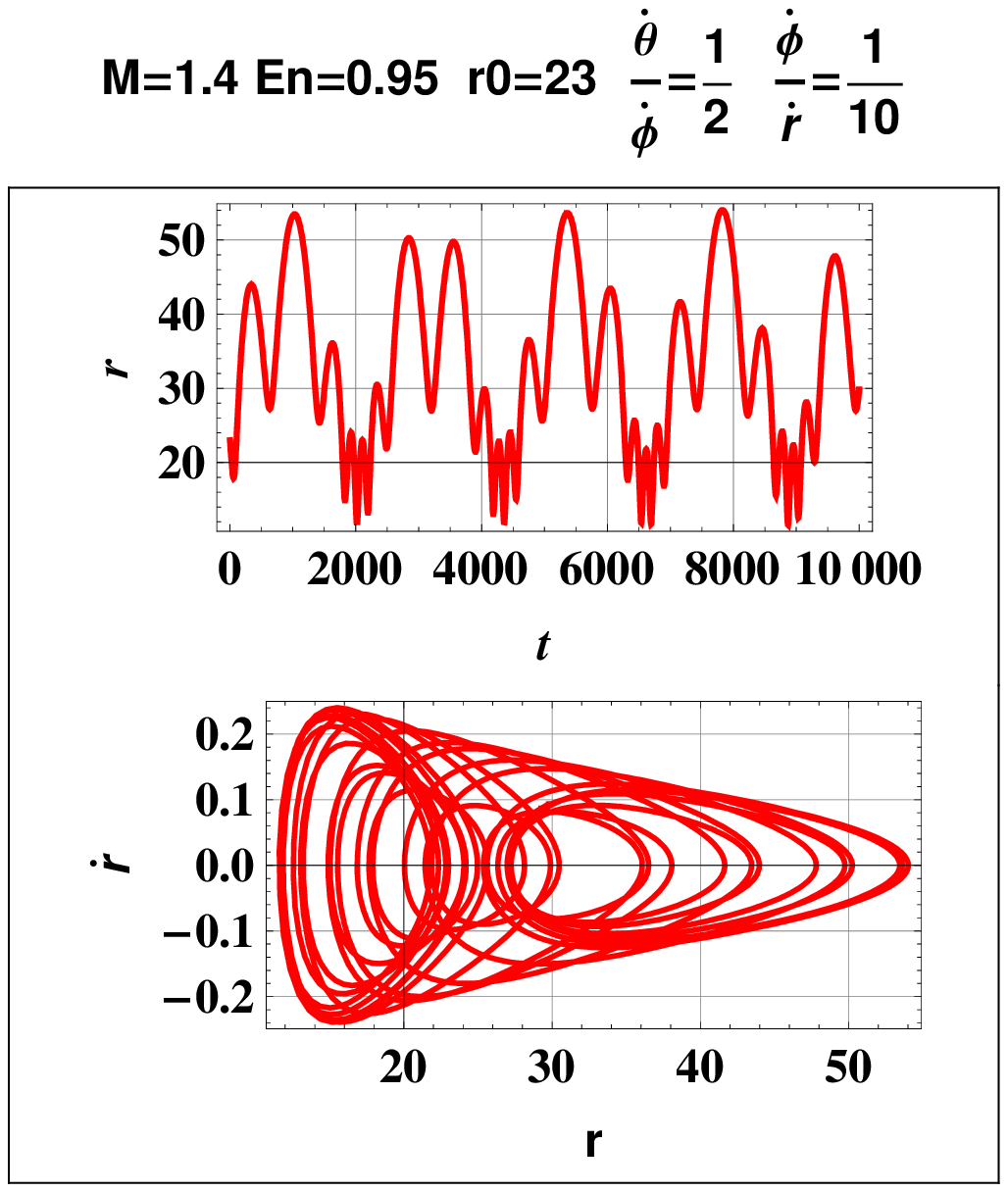}\tabularnewline
\hline
\includegraphics[scale=0.5]{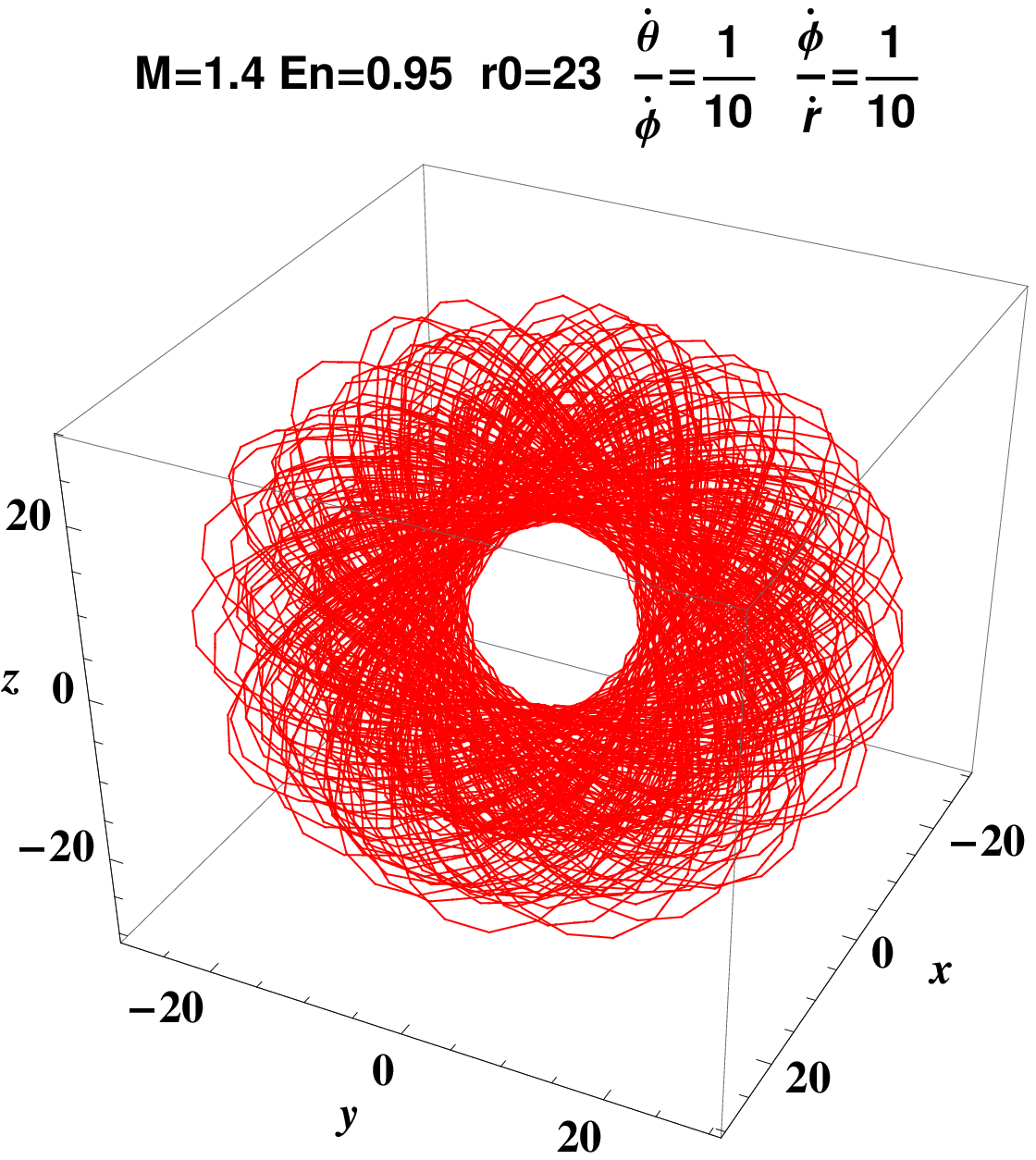} &
\includegraphics[scale=0.5]{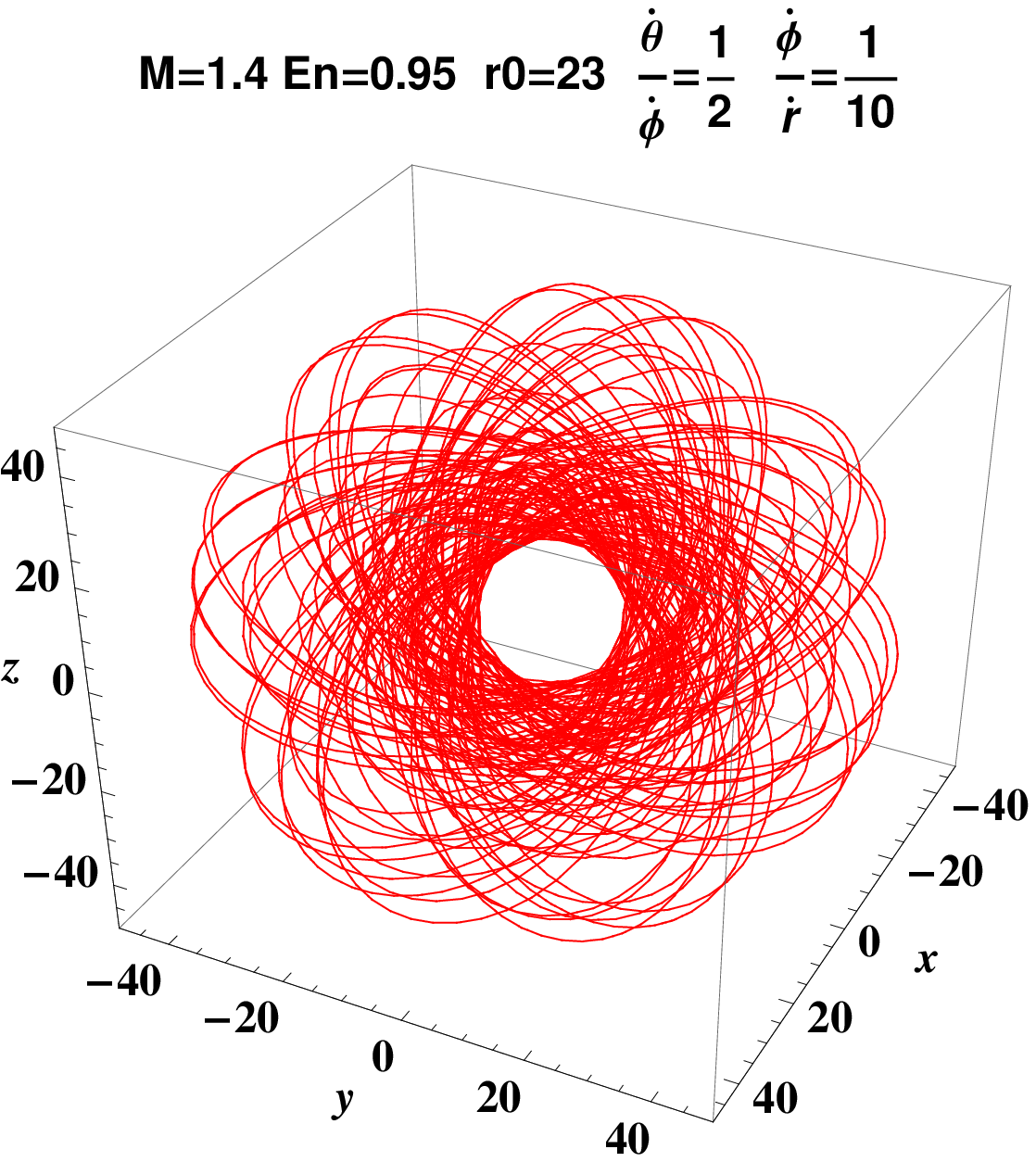}\tabularnewline
\hline
\end{tabular}
\caption {Time series $r=r(t)$ and phase portrait $\dot{r}=f(r)$
(top panels),  time series $\dot{r}=r(t)$ and $\ddot{r}=r(t)$
(middle panels), with the 3D orbits (bottom  panels) for a
Chandrasekhar mass $M=1.4$ in solar units. We assumed the
following initial conditions: $\dot{r_{0}}=-1/10$,
$\dot{\phi_{0}}=-\dot{r}_{0}/{10}$ while we have performed two
trials assuming, for the initial condition on the nutation angular
velocity $\dot{\theta}_{0}$, two limit values which we have found,
according to our empirical procedure, i.e.
$\dot{\theta}_{0}={\dot{\phi}_{0}}/{20}$ and
$\dot{\theta}_{0}={\dot{\phi_{0}}}/{2}$ respectively. At the
bottom,  the 3D orbits are plotted (left panel with
$\dot{\theta}_{0}={\dot{\phi}_{0}}/{10}$ and  the right panel with
$\dot{\theta}_{0}={\dot{\phi}_{0}}/{2}$.)} \label{fig:new}
\end{figure}

We have used for the computation the $6^{th}$ version  of {\it
Wolfram Software Mathematica} package \cite{wolfram}. The
stiffness of the differential  equations is evident from Fig.
\ref{Fig:stiffness}, where the first and second derivative of $r$,
plotted with respect to $t$, show steep peaks corresponding to the
points where the radial velocity changes its sign abruptly. We
show  the time series of both $\dot{r}(t)$ and $r(t)$ together
with the phase portrait $\dot{r}=f(r)$  and $\ddot{r}(t)$,
assuming given initial values for the angular precession and
nutation velocities (see also Fig. \ref{fig:new}. In
Fig.\ref{Fig:stiffness}, the results for a given value of nutation
angular velocity with a time span of $10000$ steps is shown. It is
interesting to see that, by increasing the initial nutation
angular velocity, being fixed all the other initial conditions, we
get curves with decreasing frequencies for $\dot{r}(t)$ and
$\ddot{r}(t)$. This fact is relevant to have an insight on  the
orbital motion stability (see Fig.\ref{fig:1}). We have taken into
account the effect of gravitomagnetic terms, in Fig.
\ref{fig:orbit}, showing the basic orbits  (left) and the orbit
with the associated velocity field in false colors (right). From a
rapid inspection of the right panel, it is clear the sudden
changes of velocity direction induced by the gravitomagnetic
effects.

To show  the orbital velocity field,  we have performed a rotation
and a projection of the orbits along the axes of maximal energy.
In other words, by a {\it Singular Value Decomposition} of the
de-trended positions and velocities, we have selected only the
eigenvectors corresponding to the largest eigenvalues and, of
course, those representing the highest energy components (see Fig
\ref{fig:orbit}).

The  above differential equations for the parametric orbital
motion are non-linear and with time-varying coefficients. In order
to have a well-posed Cauchy problem, we have to  define:
\begin{itemize}
\item the initial and final boundary condition problems;
\item the stability and the dynamical equilibrium of solutions.
\end{itemize}
We can start by  solving the Cauchy problem, as in the classical
case,  for the initial condition putting $\dot{r}=0$ ,
$\dot{\phi}=0$, $\dot{\theta}=0$ and $\theta=\frac{\pi}{2}$ and
the result we get is that the orbit is not planar being
$\ddot{\theta}\neq 0$. In this case,  we are compelled to solve
numerically the system of second order differential equations and
to treat carefully the initial conditions, taking into account the
high non-linearity of the system. A similar discussion, but for
different problems, can be found in \cite{cutler,cutler1}.

A series of numerical trials on the orbital parameters can be done
in order to get an empirical insight on the orbit stability. The
parameters involved in this analysis are the mass, the energy, the
orbital radius, the initial values of $r,\phi,\theta$ and the
angular precession and nutation velocities $\dot{\phi}$ and
$\dot{\theta}$ respectively. We have empirically assumed initial
conditions on $\dot r$, $\dot\phi$ and $\dot\theta$.

The trials we have performed can be organized in two series, i.e.
constant mass and energy variation and constant energy and mass
variation.
\begin{itemize}
\item In the first class of trials, we assume the mass equal
to $M=1M_{\bigodot}$  and the energy $E_n$ (in mass units) varying
step by step. The initial orbital radius $r_0$ can  be  changed,
according to the step in energy: this allow  to find out
numerically the dynamical equilibrium of the orbit. We have also
chosen, as varying parameters, the ratios of the precession
angular velocity $\dot\phi$ to the radial angular velocity $\dot
r$ and the ratio of the nutation angular velocity $\dot\theta$ and
the precession angular velocity $\dot\phi$. The initial condition
on $\phi$ has been assumed to be $\phi_0=0$ and the initial
condition on $\theta$ has been $\theta_0=\frac{\pi}{2}$. For $M=1$
(in Solar masses) , $\frac{\dot\theta}{\dot\phi}=\frac{1}{2}$ and
$\dot{\phi}=-\frac{\dot{r}}{10}$,  we have found out two different
empirical linear equations, according to the different values of
$\dot{\theta},\dot{\phi}$. We obtain  a rough guess of the initial
distance $r_0=r_0(E_n)$ around which is possible to find a guess
on the equilibrium of the initial radius, followed by trials and
errors procedure.

\item In the second class of trials, we have assumed the
variation of the initial orbital radius for different values of
mass at a constant energy value equal to $E_n=0.95$ in mass units.
With this conditions, we assume ${\displaystyle
\dot\phi=\frac{\dot{r}}{10}}$ and assume that $\dot{\theta}$ takes
the two values $1/2$ and $1/10$. We can approach the problem also
considering the mass parameterization, at a given fixed energy, to
have an insight of the effect of  mass variation on the initial
conditions. The masses have been varied between 0.5 and 20 Solar
masses and the distances have been found to vary according to the
two 3rd-order polynomial functions, according to the different
values of $\dot{\theta}$ with respect to the mass.
\end{itemize}

In summary, the numerical calculations, if optimized, allow to put
in evidence the specific contributions of gravitomagnetic
corrections on orbital motion. In particular, specific
contributions due to nutation and precession emerge when higher
order terms in $v/c$ are considered.

\section{Discussions and conclusions}

In this paper, we have discussed the theory of orbits considering
gravitomagnetic effects in the geodesic motion. In particular, we
have considered the orbital effects of higher-order terms in $v/c$
which is the main difference with respect to the standard approach
to the gravitomagnetism. Such terms are often discarded but, as we
have shown, they could give rise to interesting phenomena in tight
binding systems as binary systems of evolved objects (neutron
stars or black holes). They could be important for objects falling
toward extremely massive black holes as those seated in the
galactic centers \cite{cutler,cutler1}. The leading parameter for
such correction is the ratio $v/c$ which, in several physical
cases cannot be simply discarded. For a detailed discussion see
for example \cite{capozzlamb,capozzre,sereno,sereno1}. A part the
standard periastron precession effects, such terms induce
nutations and are capable of affecting the stability basin of the
orbital phase space. As shown, the global structure of such a
basin is extremely sensitive to the initial angular velocities,
the initial energy and  mass conditions which can determine
possible transitions to chaotic behaviors. Detailed studies on the
transition to chaos could greatly aid in gravitational wave
detections in order to determine the shape, the spectrum and the
intensity of the waves (for  a discussion see \cite{levin,gair}).

In a forthcoming paper, we will discuss how gravitomagnetic
effects could affect also the gravitational wave production in
extreme gravitational field regimes.

\begin{figure}[!ht]
\begin{tabular}{|c|c|}
\hline
\includegraphics[height=0.2\textheight]{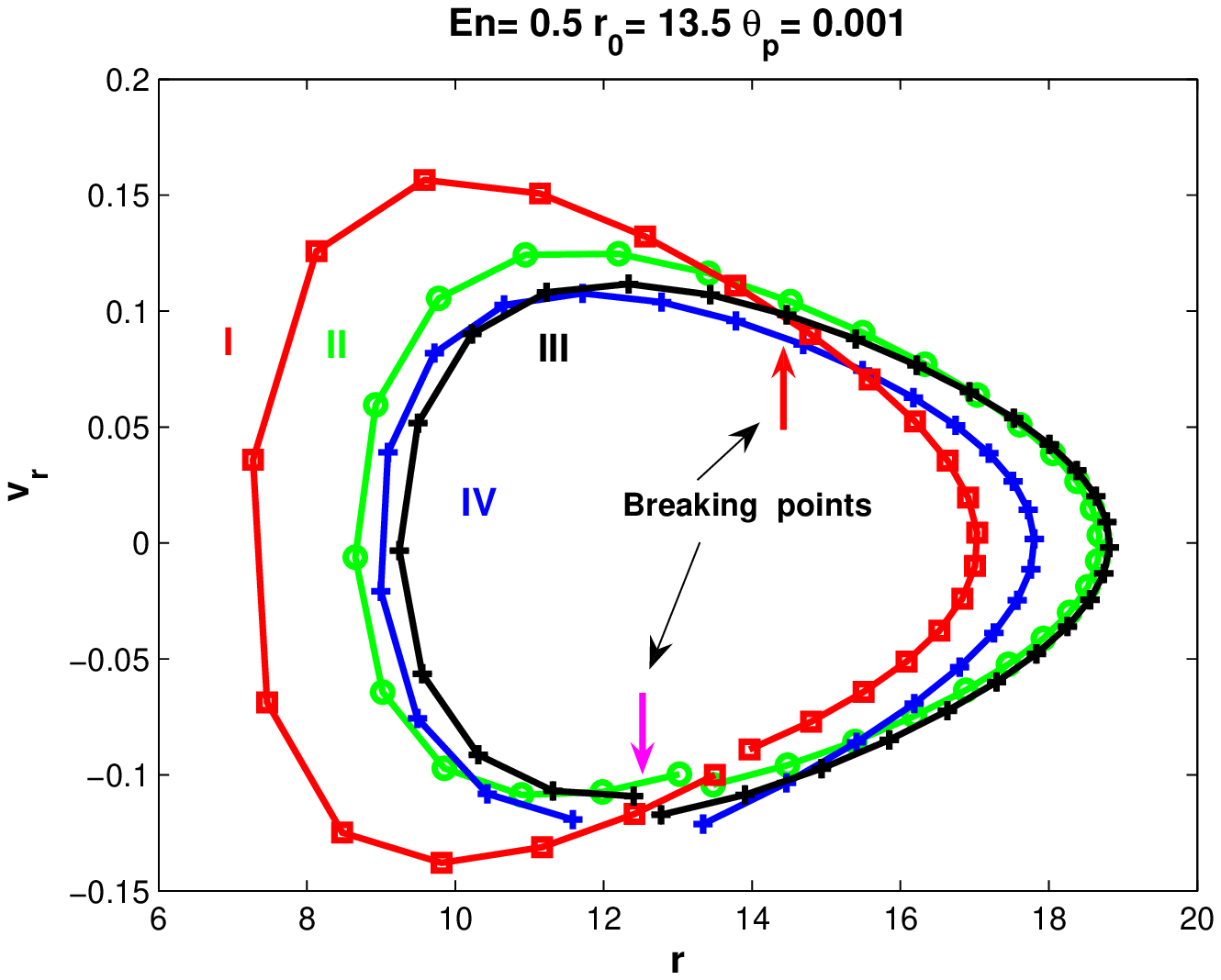}&
\includegraphics[height=0.2\textheight]{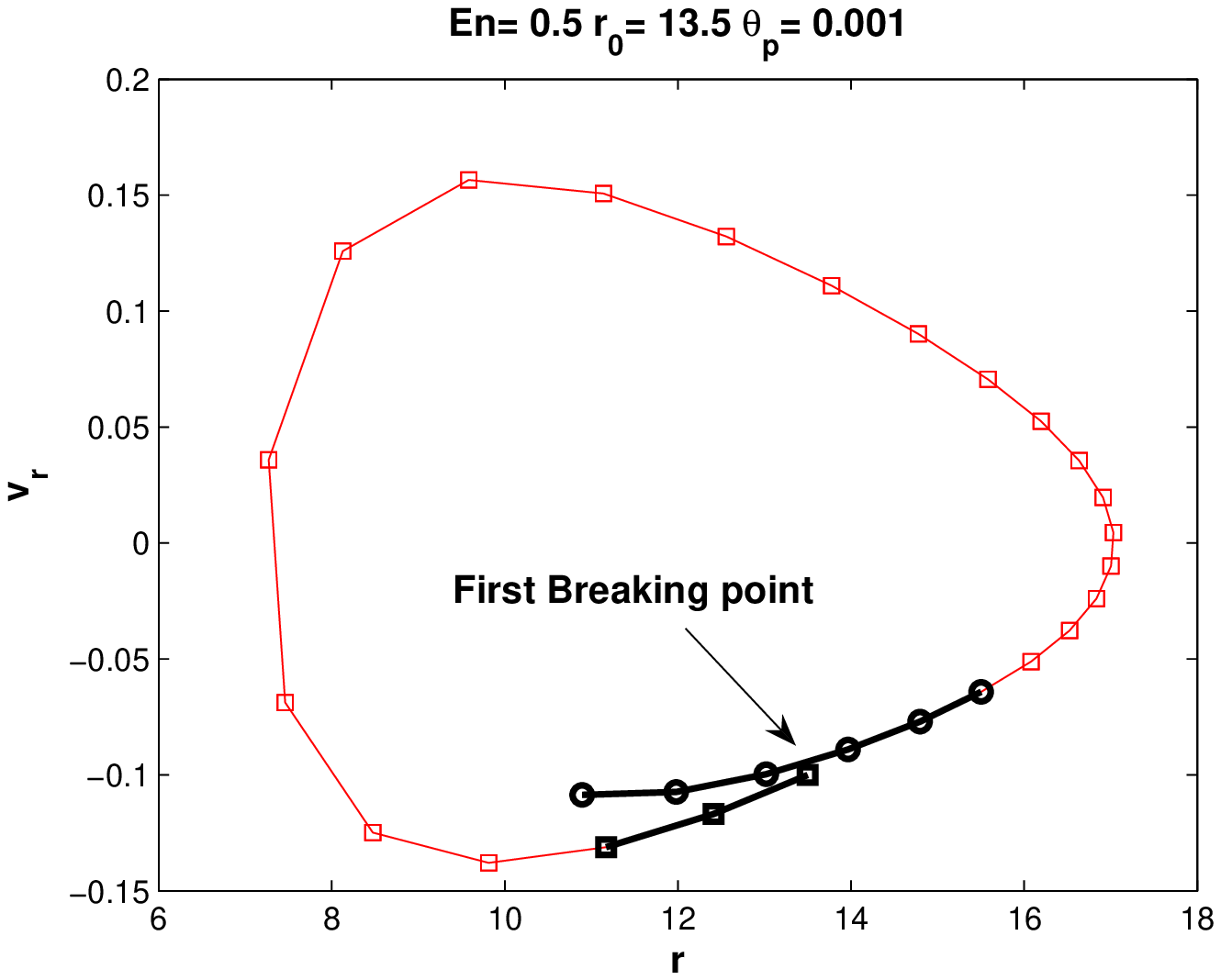}\tabularnewline
\hline
\end{tabular}
\caption {Breaking points  examples: on the left  panel, the first
four orbits in the phase plane are shown: the red one is labelled
 I, the green is II, the black is  III and the fourth is
IV. As it is possible to see, the orbits in the phase plane are
not closed and they do not overlap at the orbital closure points;
 we have called this feature {\it breaking points}. In this dynamical
situation, a small perturbation can lead the system to a
transition to the chaos as prescribed by the
Kolmogorov-Arnold-Moser (KAM) theorem  \cite{binney}. On the right
panel, it is shown the initial orbit with the initial (square) and
final (circles) points marked in black.} \label{fig:Breakstab}
\end{figure}

\end{document}